\documentstyle[12pt]{article}
\renewcommand{\baselinestretch}{2.0}
\def\lsim{\mathrel{\rlap{\lower4pt\hbox{\hskip1pt$\sim$}}
    \raise1pt\hbox{$<$}}}         
\def\gsim{\mathrel{\rlap{\lower4pt\hbox{\hskip1pt$\sim$}}
    \raise1pt\hbox{$>$}}}         
\def\ut#1{$\underline{\smash{\vphantom{y}\hbox{#1}}}$}
\def\mapright#1{\smash{
\mathop{\to}\limits_{#1}}}
\begin{document}
\noindent
THE SOLAR NEUTRINO PROBLEM\\

\noindent
\ut{W.C. Haxton}

\noindent
Institute for Nuclear Theory, NK-12, and Dept. of Physics, FM-15,
\\
University of Washington, Seattle, WA  98195\\
Haxton@phys.washington.edu\\

\noindent
KEY WORDS:  solar neutrinos, solar models, neutrino oscillations,
            neutrino detection\\

\noindent
\ut{Abstract}

\noindent
The solar neutrino problem has persisted for almost three
decades.
Recent results from Kamiokande, SAGE, and GALLEX indicate a
pattern of neutrino fluxes that is very difficult to reconcile
with plausible variations in standard solar models.  This
situation is reviewed and suggested particle physics
solutions are discussed.  A summary is given of the important
physics expected from SNO, SuperKamiokande, and other future
experiments.\\

\noindent
To Appear in Annual Reviews of Astronomy and Astrophysics 1995
\pagebreak

\noindent
CONTENTS

\noindent
1. INTRODUCTION

\noindent
2. SOLAR MODELS

     \ut{2.1 The Standard Solar Model}

     \ut{2.2 Uncertainties in Standard Model Parameters}

     \ut{2.3 Standard Model Neutrino Fluxes}

     \ut{2.4 Nonstandard Solar Models}

\noindent
3. THE DETECTION OF SOLAR NEUTRINOS

     \ut{3.1 The Homestake Experiment}

     \ut{3.2 The Kamiokande Experiment}

     \ut{3.3 The SAGE and GALLEX Experiments}

     \ut{3.4 The Molybdenum Experiment}

\noindent
4. PARTICLE PHYSICS SOLUTIONS

     \ut{4.1 Neutrino Masses and Vacuum Oscillations}

     \ut{4.2 The Mikheyev-Smirnov-Wolfenstein Mechanism}

     \ut{4.3 Other Particle Physics Scenarios}

\noindent
5. NEW EXPERIMENTS

     \ut{5.1 The Sudbury Neutrino Observatory}

     \ut{5.2 Superkamiokande}

     \ut{5.3 Other Future Detectors}

\noindent
6. OUTLOOK

\pagebreak

\noindent
1. INTRODUCTION

     Thirty years ago, in the summer of 1965, the Homestake
Mining
Company completed the excavation of the 30 x 60 x 32 ft cavity
that
was to house the 100,000-gallon chlorine detector proposed by
Ray Davis Jr. and his Brookhaven National Laboratory
collaborators.
Three years later Davis, Harmer, and Hoffman (1968) announced
the results of their first two detector runs, an upper bound on
the solar neutrino flux of 3 SNU (1 SNU = 10$^{-36}$ captures/
target atom/
sec).  The accompanying theoretical paper by Bahcall, Bahcall,
and
Shaviv (1968) found a rate of 7.5 $\pm$ 3.3 SNU for the standard
solar model.
This discrepancy persists today, augmented by nearly three
decades
of data from the Homestake experiment and by the new results from
the Kamiokande and GALLEX/SAGE detectors.  The purpose of
this review is to summarize the current status of the solar
neutrino
problem and its possible implications for physics and
astrophysics.

     A remarkable aspect of the solar neutrino problem is that it
has both endured and deepened.  The prospect of quantitatively
testing
the theory of main-sequence stellar evolution provided much of
the
original motivation for measuring solar neutrinos: solar
neutrinos
carry, in their energy distribution and flux, a precise record of
the thermonuclear reactions occurring in the sun's core.  Our
understanding of the atomic and nuclear microphysics governing
stellar evolution - nuclear reaction rates, radiative opacities,
and the equation of state - has progressed significantly since
the 1960s.  The development of helioseismology has provided a
new tool for probing the solar interior.  Finally, we better
understand our sun in the context of other stars, the
observations
of which have helped to define the envelope of possibilities
for diffusion, mass loss, magnetic fields, etc.  This progress
has tended to increase our confidence in the standard solar
model.  At the same time, with the new results from Kamiokande
and the gallium detectors, a pattern of neutrino fluxes has
emerged that is more difficult to reconcile with possible
variations in that model.

     The solar neutrino problem has deepened because of the
discovery of the Mikheyev-Smirnov-Wolfenstein (MSW) mechanism:
the sun has the potential to greatly enhance
the effects of neutrino mixing.  If neutrino oscillations prove
to be the solution to the solar neutrino problem, this will
force modifications in the standard model of electroweak
interactions, which accommodates neither massive neutrinos nor
their mixing.  This new physics would have implications for
a variety of problems in astrophysics, including the missing
mass puzzle and the formation of large-scale structure.

     There are several sources that the interested reader
can consult for additional information.  The most comprehensive
treatment is that given by Bahcall in his book ``Neutrino
Astrophysics" (1989).  The appendix of this book reprints a
delightful historical perspective of the development of the
solar neutrino problem (Bahcall and Davis 1982).  The review
by Bowles and Gavrin (1993) provides an excellent discussion
of the Homestake, Kamiokande, and SAGE/GALLEX experiments, as
well as detectors now under construction or development.
The proceedings of the Seattle Solar Modeling Workshop
(Balantekin and Bahcall 1995) and the Homestake Conference
(Cherry, Lande,
and Fowler 1985) contain many studies of the standard and
nonstandard solar models and of the nuclear and atomic
physics on which such models depend.  The theory of
matter-enhanced neutrino oscillations has been reviewed
recently by Mikheyev and Smirnov (1989).

\noindent
2. SOLAR MODELS\\
\noindent
\ut{2.1 The Standard Solar Model}

     Solar models trace the evolution of the sun over the past
4.6 billion years of main sequence burning, thereby predicting
the present-day temperature and composition profiles of the solar
core that govern neutrino production.  Standard solar models
share four basic assumptions:

 * The sun evolves in hydrostatic equilibrium, maintaining
a local balance between the gravitational force and the pressure
gradient.  To describe this condition in detail, one must
specify the equation of state as a function of temperature,
density, and composition.

     * Energy is transported by radiation and convection.  While
the solar envelope is convective, radiative transport dominates
in the core region where thermonuclear reactions take place.
The opacity depends sensitively on the solar composition,
particularly
the abundances of heavier elements.

     * Thermonuclear reaction chains generate solar energy.
The standard model predicts that over 98\% of this energy
is produced from the pp chain conversion of four protons into
$^4$He
(see Figure 1)
\begin{equation}
          4p \to ^4\textstyle{He} + 2e^+  + 2 \nu_e
\end{equation}
with the CNO cycle contributing the remaining 2\%.  The sun is
a large but slow reactor: the core temperature, $T_c \sim  1.5
\cdot 10^7$ K,
results in typical center-of-mass energies for reacting particles
of $\sim$ 10 keV, much less than the Coulomb barriers inhibiting
charged particle nuclear reactions.  Thus reaction cross
sections are small, and one must go to significantly higher
energies before laboratory measurements are feasible.  These
laboratory data must then be extrapolated to the solar energies
of
interest.

 * The model is constrained to produce today's solar
radius, mass, and luminosity.  An important assumption of
the standard model is that the sun was highly convective,
and therefore uniform in composition, when it first
entered the main sequence.  It is furthermore assumed
that the surface abundances of metals (nuclei with A $>$ 5)
were undisturbed by the subsequent evolution, and thus
provide a record of the initial solar metallicity.  The
remaining parameter is the initial $^4$He/H ratio, which
is adjusted until the model reproduces the present solar
luminosity after 4.6 billion years of evolution.  The resulting
$^4$He/H mass fraction ratio is typically 0.27 $\pm$ 0.01,
which can be compared to the big-bang value of 0.23  $\pm$ 0.01
(Walker et
al. 1991).  Note that the sun was formed from previously
processed
material.

     The model that emerges is an evolving sun.  As the
core's chemical composition changes, the opacity and
core temperature rise, producing a 44\% luminosity increase
since the onset of the main sequence.  The $^8$B neutrino
flux, the most temperature-dependent component, proves to
be of relatively recent origin: the predicted flux
increases exponentially with a doubling period of about
0.9 billion years.  The equilibrium abundance and equilibration
time
for $^3$He are both sharply increasing functions of the distance
from the solar center.  Thus a steep $^3$He density gradient is
established over time.

     The principal neutrino-producing reactions of the pp chain
and CNO cycle are summarized in Table 1.  The first six reactions
produce $\beta$ decay neutrino spectra having allowed shapes with
endpoints given by E$_\nu^{\rm max}$  .  Deviations from an
allowed spectrum
occur for $^8$B neutrinos because the $^8$Be final state is a
broad
resonance (Bahcall and Holstein 1986); much smaller deviations
occur because of
second-forbidden contributions to the decay.  The last two
reactions
produce line sources of electron capture neutrinos, with
widths $\sim$ 2 keV characteristic of the solar core temperature
(Bahcall 1993).
The resulting solar neutrino spectrum is shown in Figure 2.

   Measurements of the  pp, $^7$Be, and $^8$B neutrino fluxes
will
determine the relative contributions of the  ppI,  ppII, and
ppIII cycles to solar energy generation.  As the discusion
below will illustrate, this competition is governed in large
classes of solar models by a single parameter, the central
temperature $T_c$.  The flux predictions of two standard models,
those of Bahcall and Pinsonneault (1992) and of Turck-Chi\`eze
and Lopez (1993),
are included in Table 1.

\noindent
\ut{2.2 Uncertainties in Standard Solar Model Parameters}

     Careful analyses of the experiments that will be described
in Section 3 indicate that the observed solar neutrino fluxes
differ substantially from standard solar model (SSM) expectations
(White, Krauss, and Gates 1993; Parke 1995; Hata and Langacker
1994):
\begin{eqnarray}
\phi (pp) & \sim & 0.9 \, \phi^{\rm {SSM}} (pp)\nonumber \\
\phi (^7{\rm {Be}}) & \sim & 0 \nonumber\\
\phi (^8 {\rm B}) & \sim & 0.43 \, \phi^{\rm {SSM}} (^8{\rm B}).
\end{eqnarray}
Reduced $^7$Be and $^8$B neutrino fluxes can be produced by
lowering
the central temperature of the sun somewhat.  However, such
adjustments, either by varying the parameters of the SSM or by
adopting some nonstandard physics, tend to push the $\phi
(^7$Be)/$\phi(^8$B)
ratio to higher values rather than the low one of Eq. (2),
\begin{equation}
{\phi (^7{\rm{Be}}) \over \phi(^8 {\rm B})} \sim T_c^{-10}.
\end{equation}
Thus the observations seem difficult to reconcile with plausible
solar model variations.

  It is apparent that the rigor of this argument is the
crucial issue: how quantitative is the tracking of fluxes and
flux ratios with $T_c$, what variations can exist in models that
produce the same $T_c$ but differ in other respects, and how
significant are the results in Eq. (2) when the statistical
and systematic errors of the experiments are taken into
account?  These questions have motivated a number of careful
examinations (and clever presentations) of solar model
uncertainties.

     SSM uncertainties include the reaction cross sections for
the  pp chain and CNO cycle, the opacities, the deduction of
heavy element abundances from solar surface abundances, the
solar age and present day luminosity, and the equation of state.
Modelers occasionally adopt different ``best values" and
associated
errors for these parameters.  Bahcall and Pinsonneault (1992)
have argued that the differences among solar models are
almost always attributable to parameter choices, and not to
disagreements about the underlying physics.

     While a detailed summary of standard model uncertainties
would take us well beyond the limits of this review, a
qualitative discussion of  pp chain uncertainties is appropriate.
This nuclear microphysics has been the focus of a great deal
of experimental work, as well as the source of some contention
among modelers.  The  pp chain involves a series of nonresonant
charged-particle reactions occurring at center-of-mass energies
that are well below the height of the inhibiting Coulomb
barriers.  As the resulting small cross sections preclude
laboratory measurements at the relevant energies, one must
extrapolate higher energy measurements to threshold to
obtain solar cross sections.  This extrapolation is often
discussed in terms of the astrophysical S-factor (Fowler 1984;
Burbidge, Burbidge, Fowler, and Hoyle 1957),
\begin{equation}
\sigma (E) = {S(E) \over E} \exp (-2 \pi \eta)
\end{equation}
where $\eta = {Z_1Z_2 \alpha \over \beta}$, with $\alpha$ the
fine structure
constant and $\beta = v/c$ the relative velocity of the colliding
particles.
This
parameterization removes the gross Coulomb effects associated
with the s-wave interactions of charged, point-like particles.
The remaining energy dependence of S(E) is gentle and
can be expressed as a low-order polynomial in E.  Usually
the variation of S(E) with E is taken from a direct reaction
model and then used to extrapolate higher energy measurements
to threshold.  The model accounts for finite nuclear size
effects, strong interaction effects, contributions from
other partial waves, etc.  As laboratory measurements are
made with atomic nuclei while conditions in the solar core
guarantee the complete ionization of light nuclei, additional
corrections must be made to account for the different
electronic screening environments.

     Among the pp chain reactions, the one presently most
controversial is $^7$Be(p,$\gamma$)$^8$B.  The two experiments
performed at
the lowest energies (Kavanagh et al. 1969; Filippone et al. 1983)
find very similar energy dependences but disagree in overall
normalization by about 25\%.  A similar normalization
disagreement
exists among higher energy data sets (see Langanke 1995;
Parker 1968; Kavanagh et al. 1969; Vaughn et al. 1970).

     Theory predicts a threshold energy dependence for S(E)
that rises gently as E $\to$ 0; at sufficiently low energies,
when
the capture occurs at nuclear separations well outside the
range of nuclear forces, the behavior is fixed (Christy and
Duck 1961; Williams and Koonin 1981).  The data are consistent
with theory, though their accuracy at low energies (120-200 keV)
is insufficient to independently determine the rise at low E
(see Figure 3).  When the theoretical energy dependence (Johnson
et al. 1992) of
S(E) is used to extrapolate the Kavanagh et al. and Filippone
et al. data, one finds S(0) $\sim$ 25 eV barns and $\sim$ 20 eV
barns,
respectively.  While most calculations have predicted values
for S(0) closer to the Kavanagh value, lower values have also
been obtained (see, e.g., Xu et al. 1994) and general arguments
have been given that the theoretical normalization uncertainty
could be as much as a factor of two (Barker 1980).
Thus it does not appear that theory can distinguish between
competing experimental normalizations for S(0).  In the most
recent reexamination of the data, Johnson et al. (1992)
determined S$_{17}$(0) = 22.4  $\pm$ 2.1 eV barns, from a
weighted
average of six data sets.

     Very recently Motobayashi et al. (1994), by measuring the
breakup of $^8$B in the Coulomb field of $^{208}$Pb, deduced a
preliminary
value of S$_{17}$(0) = 16.7  $\pm$ 3.2 eV barns, which would
favor the
Filippone result.  It has been argued that the extraction of
S$_{17}$(0) from the Coulomb breakup cross section is complicated
by the three-body Coulomb effects in the outgoing channel and
by larger E2 contributions (Langanke and Shoppa 1994), though
this claim has been disputed (Gai and Bertulani 1994).
(Transverse electric multipoles of the electromagnetic current
operator of rank J are denoted EJ.)

The SSM calculations of Bahcall and Ulrich (1988) and
Turck-Chi\`eze et al. (1988) adopted S$_{17}$(0) values of 24.3
$\pm$ 1.8
eV b and 21  $\pm$ 3 eV b, respectively.  This was the most
important contribution to the differences in their flux
predictions.  Their recently updated calculations (Bahcall and
Pinsonneault 1992; Turck-Chi\`eze and Lopez 1993) both adopt
the Johnson et al. central value, though Turck-Chi\`eze and Lopez
assign a larger error bar to this result because of the concern
that the weighted average of Johnson et al. does not fully
reflect
the systematic disagreement illustrated in Figure 3.

  While there has been some movement to lower values of
S$_{17}$(0), neither the best value nor appropriate error is
likely
to be decided without additional measurements.  The resulting
reduction in $\phi$($^8$B) because of the weaker branch to the
ppIII cycle reduces the discrepancy between the SSM predictions
and the $^{37}$Cl results.  This reduction does not resolve the
solar neutrino
problem because it also leads to an increase in the predicted
ratio $\phi(^7$Be)/$\phi(^8$B), exacerbating the most puzzling
aspect
of Eq. (2).

     Small differences between the Bahcall and Pinsonneault (BP)
and Turck-Chi\`eze and Lopez (TCL) calculations also exist in
the low-energy extrapolation of the $^3$He-$^3$He data of
Krauss et al. (1987).  BP adopt the value of Parker and Rolfs
(1991), who correct for small electron screening contributions
that are believed to enhance the laboratory cross sections
obtained at the lowest energies.  The result is
S$_{33}$(0) = 5.0  $\pm$ 0.3 MeV b.  TCL adopted the central
value suggested
by Assenbaum et al. (1987) of 5.24 MeV b.  The earlier
Bahcall and Ulrich (1988) and Turck-Chi\`eze et al. calculations
had used central values of 5.15 and 5.57 MeV b, respectively,
the latter from a direct extrapolation of the Krauss et al.
data.  The somewhat lower values adopted in the most recent
calculations strengthen the branches to the  ppII and  ppIII
cycles.

\noindent
\ut{2.3 Standard Solar Model Neutrino Fluxes}

     There are some additional minor differences in the BP
and TCL SSMs, e.g., in the solar lifetime, in S$_{34}$(0), in
the treatment of plasma effects on Thompson scattering,
and in the composition.  In addition, the BP calculation
differs from TCL, and from the earlier Bahcall and Ulrich (BU)
SSM, by the inclusion of helium diffusion, which now is
the largest contributor to the differences in the resulting
flux predictions.  The fluxes for the TCL
and BP (with and without helium diffusion) SSMs are given
in Table 1.  It is reassuring that equivalent calculations
(TCL and BP without helium diffusion) differ by less than
10\% in their predictions of the temperature-dependent
$\phi(^8$B).
This flux is increased by 12\% when helium diffusion is
included.

     In the remainder of this review, citations to the BP SSM
will refer to their model with helium diffusion.

     More important than the ``best values" of the fluxes are
the ranges that can be achieved by varying the parameters of
the SSM within plausible bounds.  In order to take into account
the correlations among the fluxes when input parameters are
varied, BU constructed 1000 SSMs by
randomly varying five input parameters, the primordial
heavy-element-to-hydrogen ratio Z/X and S(0) for the  p-p,
$^3$He-$^3$He, $^3$He-$^4$He, and p-$^7$Be reactions, assuming
for
each parameter a normal distribution with the mean and standard
deviation used in their 1988 study.  (These were the parameters
assigned the largest uncertainties.)  Smaller uncertainties
from radiative opacities, the solar luminosity, and the solar
age were folded into the results of the model calculations
perturbatively, using the partial derivatives of the BU SSM
(Bahcall and
Haxton 1989).

     The resulting pattern of $^7$Be and $^8$B flux predictions
is shown in Figure 4.
The elongated error ellipses indicate that the fluxes are
strongly correlated.  Those variations producing $\phi(^8$B)
below 0.8$\phi^{\rm{SSM}}(^8$B) tend to produce a reduced
$\phi(^7$Be), but the
reduction is always less than 0.8.  Thus a greatly reduced
$\phi(^7$Be) cannot be achieved within the uncertainties assigned
to parameters in the SSM.

 A similar exploration, but including parameter variations very
far from their preferred values, was carried out by Castellani,
Degl'Innocenti, Fiorentini, Ricci, and collaborators (1994 and
1995),
who displayed their results as a function of the resulting core
temperature
$T_c$.  The pattern that emerges is striking (see Figure 5):
parameter variations producing the same value of
$T_c$ produce
remarkably similar fluxes.  Thus
$T_c$ provides an excellent
one-parameter description of standard model perturbations.
Figure 5 also illustrates the difficulty of producing a
low ratio of $\phi(^7$Be)/$\phi(^8$B) when
$T_c$ is reduced.

 The BU 1000-solar-model variations were made under the
constraint of reproducing the solar luminosity.  Those variations
show
 a similar strong correlation with $T_c$
\begin{equation}
\phi(pp) \propto T_c^{-1.2} ~~~~~~~  \phi(^7{\rm {Be}}) \propto
T_c^8 ~~~~~~~
 \phi(^8 {\rm B}) \propto T_c^{18}.
\end{equation}
Figures 4 and 5 are a compelling argument that reasonable
variations in the parameters of the SSM, or nonstandard
changes in quantities like the metallicity, opacities, or
solar age, cannot produce the pattern of fluxes deduced
from experiment (Eq. (2)).  This would seem to limit
possible solutions to errors either in the underlying physics
of the SSM or in our understanding of neutrino properties.

\noindent
\ut{2.4 Nonstandard Solar Models}

     Nonstandard solar models include both variations of SSM
parameters far outside the ranges that are generally believed
to be reasonable (some examples of which are given in Figure 5),
and changes in the underlying physics of the model.  The
solar neutrino problem has been a major stimulus to models
of the second sort.  But there are other observations
that suggest nonstandard processes could take place in the sun.
The depletion of $^7$Li in our sun (by a factor of 200) and in
other solar-like stars is not understood.  In a recent summary,
Charbonnel (1995) argued that the depletion occurs on the main
sequence after $\sim 10^8$ years and increases with time.  The
SSM predicts no solar lithium depletion: the base of the
convective zone is too shallow to reach temperatures where
lithium can be burned.  While lithium depletion in lower mass
stars ($\sim 0.9 M_\odot$) can be produced by lowering the
low-temperature
opacities, the depletion in heavier stars would seem to argue
for diffusion or mixing mechanisms not yet incorporated into
the SSM (Charbonnel 1995).

     It is far from clear that the lithium depletion problem will
require changes in the SSM that will affect the deep,
neutrino-producing regions of the sun.  On the contrary,
Charbonnel argues that the observed $^{12}$C/$^{13}$C
ratios in low-mass red giants requires a diffusion coefficient
that decreases rapidly with depth.

     Many nonstandard models were constructed to produce a
reduction in
$T_c$ of about 5\%, as this would have accounted for
the low counting rate found in the Homestake experiment.
The suggestions included models with low heavy element
abundances (``low Z" models), in which one abandons the SSM
assumption that the initial heavy element abundances are those
we measure today at the sun's surface; periodically mixed
solar cores; models where hydrogen is continually mixed into
the core by turbulent diffusion (Schatzman and Maeder 1981) or by
convective
mixing (Ezer and Cameron 1988; Shaviv and Salpeter 1968);
and models where the solar core is partially supported by a
strong central magnetic field (Abraham and Iben 1971;
Bartenwerfer 1973;
Parker 1974; Ulrich 1974) or by its rapid rotation (Demarque,
Mengel, and
Sweigart 1973),
thereby relaxing the SSM assumption that hydrostatic equilibrium
is achieved only through the gas pressure gradient.  A larger
list
is given by Bahcall and Davis (1982).  To illustrate the kinds
of consequences such models have, two of these suggestions
are discussed in more detail below.

In low-Z models (Iben 1969; Bahcall and Ulrich 1971) one
postulates a
reduction in the core metallicity from Z $\sim$ 0.02 to Z $\sim$
0.002.
This lowers the core opacity (primarily because metals
are very important to free-bound electron transitions),
thus reducing
$T_c$ and weakening the  ppII and  ppIII cycles.
The attractiveness of low-Z models is due in part to the
existence of mechanisms for adding heavier elements to the
sun's surface.  These include the infall of comets and other
debris, as well as the accumulation of dust as the sun
passes through interstellar clouds.  However, the increased
radiative
energy transport in low-Z models leads to a thin convective
envelope, in contradiction to interpretations of the
5-minute solar surface oscillations (Rhodes, Ulrich, and Simon
1977;
Christensen-Dalsgaard and Gough 1980).  A low He mass fraction
also results.  Furthermore, Michaud (1977), noting that
diffusion of material from a thin convective envelope into
the interior would deplete heavy elements at the surface,
questioned whether present abundances could have accumulated
in low-Z models.  Finally, the general consistency of solar
heavy element abundances with those observed in other main
sequence stars makes the model appear contrived.

     Models in which the solar core ($\sim$ 0.2 M$_\odot$) is
intermittently mixed (Dilke and Gough 1972; Fowler 1972;
Opik 1953) break the standard model assumption of
a steady-state sun: for a period of several million years
(approximately the Kelvin-Helmholtz time for the core)
following mixing, the usual relationship between the observed
surface luminosity and rate of energy (and neutrino) production
is altered as the sun burns out of equilibrium.  Calculations
show that both the luminosity and the  $^8$B neutrino flux are
suppressed while the sun relaxes back to the steady state.

     Such models have been considered seriously because of
instabilities associated with large gradients in the $^3$He
abundance.  In the higher temperature central regions of the
sun the  pp chain reaches equilibrium quickly.  The dominant ppI
cycle production of $^3$He is controlled by the p+p reaction rate
which varies as X$^2$T$^4$, where X and T are the local H
mass fraction and temperature, respectively, while it is
destroyed by the
$^3$He+$^3$He reaction at a rate proportional to X$_3^2$
T$^{16}$.
Thus, the core $^3$He abundance X$_3 \propto T^{-6}$ and rises
steeply
with increasing radius until a point where $^3$He
equilibrium has not yet been reached in the  pp chain.  This is
approximately
the peak of the $^3$He abundance (at r $\sim$ 0.3r$_\odot$ is the
SSM):
beyond this point the abundance declines quickly.
(See Bahcall 1989, Figure 4.2.)  This profile is unstable
under finite amplitude displacements of a volume to
smaller r: the energy released by the increased $^3$He burning
at higher T can exceed the energy in the perturbation.

   Dilke and Gough (1972), in the ``solar spoon," proposed
that the increased $^3$He burning would also produce a linear
instability of low-order, low-degree gravity modes which
they postulated could trigger large-scale mixing of the solar
core.  Under small oscillations, the enhanced burning in
a volume element displaced downward increases the bouyant
restoring force, leading to greater ascending velocities than
descending ones.
     The question is whether this instability survives damping
mechanisms such as radiative diffusion, turbulent convection,
and couplings to higher order g-modes.  As Merryfield (1995)
recently summarized, the possibility of an instability is
still open, though no calculation has included all of the
damping mechanisms thought to be important.  He is far less
optimistic about the possibility that the instability would
drive large-scale, intermittent mixing (Merryfield, Toomre,
and Gough 1990).  Theory suggests that
the amplitude of such oscillations would saturate at velocities
of $\sim$ 10 km/s; but observation seems to rule out such large
amplitude, low-order, low-degree g-modes.  The expected
surface velocities of such modes would be comparable to
the core velocities, exceeding the helioseismology
bound of $\sim$ 10 cm/s substantially.

  These discussions of two of the more seriously explored
nonstandard models illustrate how changes motivated by the
solar neutrino problem often produce other, unwanted
consequences.  Both examples underscore the growing importance
of helioseismology as a test of the SSM and as a constraint on
its possible variations.

     Figure 6 is an illustration by Hata (1995) of the flux
predictions of several nonstandard models, including a low-Z
model consistent with the $^{37}$Cl results.  As in the
Castellani
et al. exploration, the results cluster along a track that
defines the naive
$T_c$ dependence of the $\phi (^7$Be)/$\phi(^8$B) ratio,
well separated from the experimental contours.

Of course, one cannot rule out a nonstandard solar model
solution to the solar neutrino problem: the lack of success
to date may merely reflect our lack of creativity.  But if
such a model exists - one that is consistent with our
general observations of main-sequence evolution of solar-type
stars and with helioseismology - it likely involves some
new and subtle physics.

\noindent
3. THE DETECTION OF SOLAR NEUTRINOS

     Four solar neutrino experiments have now provided data,
the Homestake $^{37}$Cl experiment, the gallium experiments SAGE
and
GALLEX, and Kamiokande.  The first three detectors are
radiochemical, while Kamiokande records neutrino-electron
elastic scattering event-by-event.

\noindent
\ut{3.1 The Homestake Experiment}

     Detection of neutrinos by the reaction
$^{37}$Cl($\nu_e$,e)$^{37}$Ar
was suggested independently by Pontecorvo (1946) and by
Alvarez (1949).  Davis's efforts to mount a 0.61 kiloton
experiment using perchloroethylene (C$_2$Cl$_{4}$) were greatly
helped
by Bahcall's demonstration (1964a, 1964b) that transitions to
excited states in $^{37}$Ar, particularly the superallowed
transition to the analog state at 4.99 MeV, increased the
 $^8$B cross section by a factor of 40.  This suggested that
Davis's detector would have the requisite sensitivity to
detect  $^8$B neutrinos, thereby accurately determining the
central temperature of the sun.  The experiment (see Figure 7)
was mounted
in the Homestake Gold Mine, Lead, South Dakota, in a cavity
constructed approximately 4850 feet underground [4900
meters water equivalent (m.w.e.)].  It has operated
continuously since 1967 apart from a 17 month hiatus
in 1985/86 caused by the failure of the circulation pumps.
The result of 25 years of measurement is (Lande 1995)
\begin{equation}
\langle \sigma \phi\rangle_{^{37}Cl} = 2.55 \pm 0.17 \pm 0.18 ~~
{\rm {SNU}}
{}~~~~(1 \sigma)
\end{equation}
which can be compared to the BP and
TCL SSM predictions
of 8.0  $\pm$ 1.0 SNU and 6.4  $\pm$ 1.4 SNU, respectively,
all with 1$\sigma$ errors.  As we will discuss below, the
 $^8$B and $^7$Be contributions account for 77\% (73\%) and
15\% (17\%), respectively, of the BP (TCL) total.

 The experiment (Davis 1985; Davis 1993)
depends on the special properties of $^{37}$Ar:
as a noble gas, it can be removed readily from perchloroethylene,
while its half life ($\tau_{1/2}$ = 35 days) allows both a
reasonable
exposure time and counting of the gas as it decays back to
$^{37}$Cl.  Argon is removed from the tank by a helium purge, and
the gas then circulated through a condensor, a molecular sieve,
and a charcoal trap cooled to the temperature of liquid nitrogen.
Typically $\sim$ 95\% of the argon in the tank is captured in the
trap.  (The efficiency is determined each run from the recovery
results for a known amount of carrier gas, $^{36}$Ar or
$^{38}$Ar,
introduced into the tank at the start of the run.)  When the
extraction is completed, the trap is heated and swept by He.
The extracted gas is passed through a hot titanium filter to
remove reactive gases, and then other noble gases are separated
by gas chromatography.  The purified argon is loaded into
a small proportional counter along with tritium-free methane,
which serves as a counting gas.  Since the electron capture
decay of $^{37}$Ar leads to the ground state of $^{37}$Cl, the
only
signal for the decay is the 2.82 keV Auger electron produced
as the atomic electrons in  $^{37}$Cl adjust to fill the K-shell
vacancy.  The counting of the gas typically continues for
about one year ($\sim$ 10 half lives).

The measured cosmic ray-induced background in the Homestake
detector is 0.06
$^{37}$Ar atoms/day while neutron-induced backgrounds are
estimated to be below
0.03 atoms/day.  A signal of 0.48  $\pm$ 0.04 atoms/day is
attributed to solar
neutrinos.  When detector efficiencies, $^{37}$Ar decays
occurring
in the tank, etc., are taken into account, the number of
 $^{37}$Ar atoms counted is about 25/year.

     A variety of careful tests of the argon recovery and
counting efficiency have been made over the past 25 years.  For
example, known amounts of $^{36}$Cl were introduced into the tank
in order to check the recovery of $^{36}$Ar, with a resulting
yield of 100   $\pm$ 3\%.  It has also been verified that
$^{37}$Ar produced in the tank by a fast neutron source,
which induces (n,p) reactions followed by (p,n) on $^{37}$Cl,
is quantitatively recovered.  However, the detector has never
been
calibrated directly with a neutrino source, despite studies of
the feasibility of a $^{65}$Zn source (Alvarez 1973).

 The significance of the Homestake results is due in part
to an accurately determined $^{37}$Cl cross section.
As the 814 keV threshold for exciting the $^{37}$Ar ground state
is above the  pp endpoint, the detector is sensitive primarily
to  $^7$Be and  $^8$B neutrinos (see Table 2).  The cross section
for  $^7$Be neutrinos
(and the weaker fluxes of pep and CNO cycle neutrinos) is
determined by the known half life of  $^{37}$Ar.  However,  $^8$B
neutrinos can generate transitions to many excited states
below the particle breakup threshold in  $^{37}$Ar.  The
superallowed
transition to the 4.99 MeV state, dominated by the Fermi matrix
element
of known strength, accounts for about 60\% of the SSM cross
section.
The allowed transition strengths
can be measured by observing the delayed protons following
the $\beta$ decay of $^{37}$Ca, the isospin analog of the
reaction
 $^{37}$Cl($\nu_e$,e) $^{37}$Ar (Bahcall 1966).  While it was
believed that
this measurement had been properly done many years ago
(Poskanzer et al. 1966; Sextro, Gough, and Cerny 1974),
the issue was not resolved until kinematically complete
measurements were done recently (Garcia et al. 1991 and 1995;
Adelberger and Haxton 1987).  The net result is a $^{37}$Cl cross
section
believed to be accurate to about 3\%.

\noindent
\ut{3.2 The Kamiokande Experiment}

     The Kamiokande experiment (Hirata et al. 1988; Hirata et al.
1991) is a 4.5 kiloton cylindrical
imaging water Cerenkov detector originally designed for proton
decay searches, but later reinstrumented to detect low energy
neutrinos.
It detects neutrinos by the Cerenkov light produced by recoiling
electrons in the reaction
\begin{equation}
\nu_x + e \to \nu'_x + e'.
\end{equation}
Both $\nu_e$ and heavy flavor neutrinos contribute, with $\sigma
(\nu_e)/\sigma(\nu_\mu) \sim$ 7.
The inner volume of 2.14 kilotons is viewed by 948 Hamamatsu 20"
photomultiplier tubes (PMTs) providing 20\% photocathode
coverage,
and the surrounding 1.5m of water, serving as an anticounter,
is viewed by 123 PMTs.  The fiducial volume for solar neutrino
measurements is the central 0.68 kilotons of water, the detector
region most isolated from the high energy gamma rays generated
in the surrounding rock walls of the Kamioka mine.

  In the conversion of the original proton decay detector
to Kamiokande II, great effort was invested in reducing low
energy backgrounds associated with radon and uranium.
This included sealing the detector against radon inleakage
and recirculating the water through ion exchange columns.
The relatively shallow depth of the Kamioka mine (2700 m.w.e.)
leads to an appreciable flux of cosmic ray muons which, on
interacting with $^{16}$O, produce various short-lived spallation
products.  These $\beta$ decay activities are vetoed by their
correlation in time with the muons.  The experimenters
succeeded in lowering the detector threshold to 9 MeV and
later to 7.5 MeV.  Kamiokande III included improvements in
the electronics and the installation of wavelength shifters
around the PMTs to increase light collection and currently
operates
with a threshold of 7.0 MeV.

   Kamiokande II/III detects the high energy portion of the
 $^8$B neutrino spectrum.  Between December, 1985, and July,
1993,
1667 live detector days of data were accumulated.  Under the
assumption that the incident neutrinos are $\nu_e$s with an
undistorted  $^8$B $\beta$ decay spectrum, the combined
Kamiokande II/III
data set gives (Nakamura 1994)
\begin{equation}
\phi_{\nu_e} (^8{\rm B}) = (2.91 \pm 0.08  \pm 0.12)\cdot
10^6/cm^2 s
{}~~~~(1\sigma)
\end{equation}
corresponding to 51\% of the BP and 63\% of the TCL SSM
predictions.
The total number of detected solar neutrino events is
476$^{+36}_{-34}$.

This experiment is remarkable in several respects.  It is
the first detector to measure solar neutrinos in real time.
Essential to the experiment is the sharp peaking of the electron
angular distribution in the direction of the incident neutrino:
this forward peaking, illustrated in Figure 8, allows the
experimenters to separate solar neutrino events from an
isotropic background.  The unambiguous observation of a peak
in the cross section correlated with the position of the sun
is the first direct demonstration that the sun produces neutrinos
as a byproduct of fusion.  Finally, although reaction (9)
is a soft process where the recoil electron and scattered
neutrino share the initial energy, the recoil electron energy
distribution provides some information on the incident neutrino
spectrum.  The recoil spectrum measured by Kamiokande II/III,
shown in Figure 9, is consistent with an allowed  $^8$B incident
neutrino spectrum, with the overall flux reduced as in Eq. (11).

\noindent
\ut{3.3 The SAGE and GALLEX Experiments}

     Two radiochemical gallium experiments exploiting the
reaction
$^{71}$Ga($\nu_e$,e)$^{71}$Ge, SAGE and GALLEX, began solar
neutrino measurements
in January, 1990, and May, 1991, respectively.  SAGE operates in
the Baksan Neutrino Observatory, under 4700 m.w.e. of shielding
from Mount Andyrchi in the Caucasus, while GALLEX is housed in
the Gran Sasso Laboratory at a depth of 3300 m.w.e.  These
experiments are sensitive primarily to the
low-energy  pp neutrinos, the flux of which is sharply
constrained
by the solar luminosity in any steady-state model of the sun (see
Table 2).
The gallium experiment was first suggested by Kuzmin (1966).
In 1974 Ray Davis and collaborators began work to develop a
practical experimental scheme.  Their efforts, in which both
GaCl$_3$ solutions and Ga metal targets were explored, culminated
with the
1.3-ton Brookhaven/Heidelberg/Rehovot/Princeton pilot experiment
in 1980-82 that demonstrated
the procedures later used by GALLEX (Hampel 1985).

 The primary obstacles to mounting the gallium experiments
were the cost of the target and the greater complexity of the
$^{71}$Ge chemical extraction.  The GALLEX experiment (Anselmann
et al.
1992 and 1994a) employs
30 tons of Ga as a solution of GaCl$_3$ in hydrochloric acid.
After an exposure of about three weeks, the Ge is recovered as
GeCl$_4$ by bubbling nitrogen through the solution and then
scrubbing the gas through a water absorber.  The Ge is further
concentrated and purified, and finally converted into GeH$_4$
which,
when mixed with Xe, makes a good counting gas.  The overall
extraction efficiency is typically 99\%.  The GeH$_4$ is inserted
into miniaturized gas proportional counters, carefully designed
for their radiopurity, and the Ge counted as it decays back to Ga
($\tau_{1/2}$ = 11.43 d).  As in the case of  $^{37}$Ar, the only
signal for
the Ge decay is the energy deposited by Auger electrons and
x-rays that accompany the atomic rearrangement in Ga.  An
important achievement of GALLEX has been the detection of both
the K peak (10.4 keV) and L peak (1.2 keV).  While 88\% of the
electron captures occur from the K shell, many of the subsequent
K $\to$  L x-rays escape the detector and some of the Auger
electrons
hit the detector walls.  This produces a shift of the detected
energy of events from the K to the L and M peaks.  Thus
the GALLEX L-peak counting capability almost doubles the
$^{71}$Ge
detection efficiency.

 Gallium, like mercury, is a liquid metal at room temperature.
SAGE (Abdurashitov 1994; Bowles and Gavrin 1993)
uses metallic gallium as a target, separating the $^{71}$Ge
by vigorously mixing into the gallium a mixture of hydrogen
peroxide and dilute hydrochloric acid.  This produces an
emulsion, with the Ge migrating to the surface of the emulsion
droplets where it is oxidized and dissolved by hydrochloric acid.
The Ge is extracted as GeCl$_4$, purified and concentrated,
synthesized into GeH$_4$, and further purified by gas
chromatography.
The overall efficiency, determined by introducing a Ge carrier,
is typically 80\%.  The Ge counting proceeds as in GALLEX.
     SAGE began operations with 30 tons of gallium, and now
operates with 55 tons.  The combined result for stage I (prior
to September, 1992) and the first nine runs of stage II (9/92 -
6/93) is (Nico 1995)
\begin{equation}
\langle \sigma \phi \rangle_{^{71}{\rm {Ge}}} = 69^{+11}_{-11}\,
{\rm
{(stat)}}^{+5}_{-7} \, {\rm {(sys)}} \,\,\, {\rm
{SNU}}~~~~(1\sigma) .
\end{equation}
This result includes only events counted in the K peak.
The counter and electronics improvements made at the start of
stage II should permit L-peak events to be included, but no
results have been announced as of December, 1994.  The
corresponding results for GALLEX I (first 15 runs) and II
(8/92-10/93) is (Anselmann et al. 1994)
\begin{equation}
\langle \sigma \phi \rangle_{^{71}{\rm {Ge}}} = 79 \pm 10 \pm \,
6 \,\,
{\rm {SNU}} ~~~~(1 \sigma) .
\end{equation}
GALLEX II solar neutrino runs continued until June, 1994,
but results for the last portion of this period have not been
reported.

GALLEX II solar neutrino runs were interrupted in June, 1994,
to permit an overall test of the detector with a
$^{51}$Cr source, which produces line sources of 746 keV (90\%)
and 426 keV (10\%) neutrinos.  The 1.67 MCi source was produced
by irradiating $\sim$ 36 kg of chromium, enriched in $^{50}$Cr,
in
the Silo\'e reactor in Grenoble.  Following exposure of the
detector and recovery and counting of the produced $^{71}$Ge,
the ratio of measured $^{71}$Ge to expected was calculated
(Anselmann et al. 1995),
\begin{equation}
              R = 1.04  \pm 0.12~~~(1\sigma).
\end{equation}
This is the first test of a solar neutrino detector with an
terrestrial, low energy
neutrino source.  A similar source ($\sim$ 0.5 MCi) has been
produced by
the SAGE collaboration and was installed in their
detector in December, 1994 (J.F. Wilkerson, private
communication).  The
higher Ga density of the SAGE detector increases the
effectiveness
of the source by about a factor of 2.5, helping to compensate for
the weaker neutrino flux.

The nuclear physics of the reaction $^{71}$Ga($\nu_e$,e)$^{71}$Ge
is
illustrated in Figure 10.  As the threshold is 233 keV, the
ground state and first excited state can be excited by  pp
neutrinos.  However,
as only those  pp neutrinos within 12 keV of the endpoint can
reach the excited
state, the phase space for reaching this state is smaller by a
factor of $\sim$ 100.
Thus the cross section is determined precisely by the measured
electron capture lifetime of $^{71}$Ge.  In the BP and TCL
calculations these neutrinos account for 54\% and 57\% of
the capture rate, respectively, each predicting 71 SNU.
Because of this strong  pp neutrino contribution, there
exists a minimal astronomical counting rate of 79 SNU
(Bahcall 1989) for the Ga detector that assumes only a
steady-state sun and standard model weak interaction physics.
This minimum value corresponds to a sun that produces the
observed luminosity entirely through the  ppI cycle.
The rates found by SAGE and GALLEX are quite close to this
bound.

The  $^7$Be neutrinos can excite the ground state and two
excited states at 175 keV (5/2$^-$) and 500 keV (3/2$^-$).  The
SSM
rates quoted above assume that these excited state transitions
are much weaker than the ground state transition, contributing
only 5\% to the  $^7$Be rate.  The primary justification for this
is the forward-angle (p,n) calibration of Gamow-Teller
transitions to these states by Krofcheck et al. (1985).
However, while the track record of (p,n) mappings of the
broad features of the Gamow-Teller resonance is quite good,
the technique is not generally considered a reliable
test of the GT strength carried by an isolated state unless
the transition to that state is quite strong (Austin et al.
1994).
On the other hand, all three  $^7$Be transitions are tested in
the $^{51}$Cr calibration, weighted by phase space factors quite
similar to those of  $^7$Be neutrinos (Haxton 1988a).  Given
other checks the GALLEX collaboration has made of the
detector's overall chemical efficiency, another interpretation
(Hata and
Haxton 1995) of the source experiment is that it verifies that
the excited
states play a minor role in the  $^7$Be capture rate.  As
the SSM $^7$Be capture rates of BP and TCL are both above 30 SNU,
the SAGE/GALLEX results alone suggest some reduction in the
low-energy pp and $^7$Be fluxes.

  The  $^8$B neutrino capture rates (14 and 11 SNU in the BP
and TCL SSMs, respectively) have been calculated from the GT
profile deduced by
Krofcheck et al. (1985).  The corresponding total SSM rates for
this detector are 132 SNU
and 123 SNU (see Table 2).

\noindent
\ut{3.4 The Molybdenum Experiment}

     Recently an effort was made to do a solar neutrino
experiment
of a different kind, a geochemical integration of the  $^8$B
neutrino
flux over the past several million years (Cowan and Haxton 1982;
Wolfsberg 1985).  The reaction
\begin{equation}
\nu_e + ^{98}{\rm {Mo}} \to e^- + ^{98}{\rm {Tc}} \,\,
(\tau_{1/2} = 4.2 \cdot 10^6
y)
\end{equation}
occurring over geologic times can produce concentrations of
$^{98}$Tc
of $\sim$ 10 atoms per gram of $^{98}$Mo (abundance 24\%).
Because no
stable isotopes of technetium exist, such a concentration is,
in principle, measureable.  Careful calculations of backgrounds
from natural radioactivity and cosmic rays indicated they
would be tolerable in a deeply buried ore body.

 The motivation of the experiment differs from others we
have discussed: a comparison of  $^8$B flux averaged over an
appreciable fraction of the Kelvin-Helmholtz time with
contemporary measurements would test the long-term thermal
stability of the core.

     The challenge is to isolate and quantitatively count $\sim
10^8$
atoms of $^{98}$Tc from 2600 tons of raw ore (containing 13 tons
of molybdenum) obtained from the one operating deep mine in
North America, the Henderson Mine in Clear Creek County,
Colorado.  Two major tasks faced the experimenters, isolating
the Tc from this enormous quantity of ore and doing quantitative
 Tc mass spectrometry at the required level.  The first goal
appeared reachable as it is known that chemically analogous
rhenium can be quantitatively recovered from the gas stream
of commercial molybdenum roasters.  The second goal was reached
after several years of work.

     The experiment was mounted by Los Alamos National Laboratory
with the help of the AMAX Mining Corporation.  The experimenters
achieved a sensitivity to Tc at about the 100 SNU level, five
times the expected SSM production rate.  At that level a
background appeared due to the commercial roaster's ``memory"
of recently roasted molybdenum from shallow mines in which
cosmic-ray-induced  Tc levels are high.  While the plant memory
decays
with time, the necessary flushing of the roaster with several
weeks of Henderson ore was not practical commercially.  The
effort was abandoned in 1988.

  It appears this experiment must await a factor of 10-100
further improvement in  Tc mass spectrometry.  At this
sensitivity,
the roasting of the MoS$_2$ concentrate could be done under
controlled
laboratory conditions on a ``table-top" scale of 10-100 kgs.

\noindent
4. PARTICLE PHYSICS SOLUTIONS

     In Section 2 it was shown that solar models which reduce
the high energy neutrino flux tend to enhance the  $^7$Be/$^8$B
flux
ratio, contradicting the results of the Homestake,
SAGE/GALLEX, and Kamiokande experiments.

  If the source of the solar neutrino problem is not
solar, the remaining possibilities are experimental error
or nonstandard particle physics.  Several researchers have
recently considered the consequences of ignoring one of
the three experiments just discussed (Parke 1995; Bahcall
1994; Kwong and Rosen, 1994; Bahcall and Bethe 1990).
Substantial discrepancies
between SSM predictions and the remaining experiments
persist.  For example, Figure 11 shows Parke's results
when only the Kamiokande and SAGE/GALLEX constraints are
retained: a discrepancy of almost 3-4 $\sigma$ remains,
depending on the choice of SSM.  The corresponding results
for other combinations show even larger inconsistencies.
If two experiments must be flawed to account for the
discrepancy with the SSM, this scenario becomes somewhat
less credible.

     The second alternative, physics beyond the standard model of
electroweak interactions, would have the most far-reaching
consequences.  Particle physics solutions of the solar
neutrino problem include neutrino oscillations, neutrino
decay, neutrino magnetic moments, and weakly interacting
massive particles.  Among these, the Mikheyev-Smirnov-Wolfenstein
effect $-$ neutrino oscillations enhanced by matter
interactions $-$ is widely regarded as the most plausible.

\noindent
\ut{4.1 Neutrino Masses and Vacuum Oscillations}

     One odd feature of particle physics is that neutrinos,
which are not required by any symmetry to be massless,
nevertheless
must be much lighter than any of the other known fermions.
For instance, the current limit on the $\overline{\nu}_e$ mass is
$\lsim$ 5 eV.
The standard model requires neutrinos to be massless, but the
reasons are not fundamental.  Dirac mass terms $m_D$, analogous
to the mass terms for other fermions, cannot be constructed
because the model contains no right-handed neutrino fields.
Neutrinos can also have Majorana mass terms
\begin{equation}
\overline{\nu^c_L} m_L \nu_L ~~~ and ~~~ \overline{\nu^c_R} m_R
\nu_R.
\end{equation}
where the subscripts $L$ and $R$ denote left- and right-handed
projections
of the neutrino field $\nu$, and the superscript $c$ denotes
charge conjugation.
The first term above is constructed from left-handed fields, but
can
only arise as a nonrenormalizable effective interaction when
one is constrained to generate $m_L$ with the doublet scalar
field of
the standard model.  The second term is absent from the standard
model because there are no right-handed neutrino fields.

     None of these standard model arguments
carries over to the more general, unified theories that
theorists believe will supplant the standard model.
In the enlarged multiplets of extended models it is
natural to characterize the fermions of a single family,
e.g., $\nu_e$, e, u, d, by the same mass scale $m_D$.  Small
neutrino
masses are then frequently explained as a result of the
Majorana neutrino masses.  In the seesaw mechanism,
\begin{equation}
M_\nu \sim \left(\begin{array}{cc}
0 & m_D \\
m^T_D & m_R \end{array}\right).
\end{equation}
Diagonalization of the mass matrix
produces one light neutrino, $m_{\rm {light}}\sim {m_D^2 \over
m_R}$, and one
unobservably
heavy, $m_{\rm {heavy}} \sim m_R$.  The factor ($m_D$/$m_R$) is
the needed small
parameter that accounts for the distinct scale of neutrino
masses.  The masses for the $\nu_e, \nu_\mu$, and $\nu_\tau$ are
then
related to the squares of the corresponding quark masses
$m_u$, $m_c$, and $m_t$.  Taking $m_R \sim 10^{16}$ GeV, a
typical grand
unification scale for models built on groups like SO(10), the
seesaw mechanism gives the crude relation
\begin{equation}
m_{\nu_e}: m_{\nu_\mu}: m_{\nu_\tau} \leftrightarrow 2 \cdot
10^{-12}: 2
\cdot 10^{-7}: 3 \cdot 10^{-3} eV.
\end{equation}
The fact that solar neutrino experiments can probe small
neutrino masses, and thus provide insight into possible new
mass scales $m_R$ that are far beyond the reach of direct
accelerator measurements, has been an important theme of
the field (Babu and Mohapatra 1993; Bludman, Kennedy, and
Langacker, 1992;
Dimopoulos, Hall, and Raby 1993).

 Another expectation is that neutrinos from the different
families mix, just as quark mixing is observed in hyperon and
nucleon $\beta$ decays.  If we consider the two-flavor case for
simplicity, the mass eigenstates $|\nu_1\rangle$ and
$|\nu_2\rangle$ (with
 masses
$m_1$ and $m_2$) are related to the weak interaction eigenstates
by
\begin{eqnarray}
|\nu_e\rangle &=& \cos \theta_v |\nu_1\rangle + \sin
\theta_v|\nu_2 \rangle
\nonumber \\
|\nu_\mu\rangle &=& - \sin \theta_v |\nu_1 \rangle + \cos
\theta_v |\nu_2
\rangle
\end{eqnarray}
where $\theta_v$ is the (vacuum) mixing angle.  The two mass
eigenstates
comprising the $\nu_e$ then propagate with different phases in
vacuum, leading to flavor oscillations.  The probability
that a $\nu_e$ will remain a $\nu_e$ after propagating a distance
$x$ is
\begin{equation}
P_{\nu_e} (x) = 1 - \sin^2 2 \theta_v \sin^2 \left({\delta m^2 x
\over 4
E}\right) \mapright{x \to \infty}\ 1 - {1 \over 2} \sin^2 2
\theta_v
\end{equation}
where E is the neutrino energy and $\delta m^2  = m^2_2 - m^2_1$.
(When one properly describes the neutrino state as a wave packet,
the large-distance behavior follows from the eventual separation
of the mass eigenstates.)  If the
the oscillation length
\begin{equation}
L_o = {4 \pi E \over \delta m^2}
\end{equation}
is comparable to or shorter than one astronomical unit, a
reduction in the $\nu_e$ flux would be expected in terrestrial
neutrino oscillations.
     The suggestion that the solar neutrino problem could
be explained by neutrino oscillations was first made by
Pontecorvo (1958), who pointed out the analogy with $K_0
\leftrightarrow \bar
K_0$
oscillations.  From the point of view of particle physics,
the sun is a marvelous neutrino source.  The neutrinos travel a
long
distance and have low energies ($\sim$ 1 MeV), implying a
sensitivity to
\begin{equation}
\delta m^2 \gsim 10^{-12} eV^2.
\end{equation}
In the seesaw mechanism, $\delta m^2 \sim m^2_2$, so neutrino
masses as
low as $m_2 \sim 10^{-6}eV$ could be probed.  In contrast,
terrestrial
oscillation experiments with accelerator or reactor
neutrinos are typically limited to $\delta m^2 \gsim 0.1 eV^2$.

     From Eq. (17) one expects vacuum oscillations to affect
all neutrino species equally, if the oscillation length is small
compared to an astronomical unit.  This appears to contradict
observation,
as the  pp flux may not be significantly reduced.
Furthermore, the theoretical prejudice that $\theta_v$ should be
small makes this an unlikely explanation of the significant
discrepancies with SSM  $^7$Be and  $^8$B flux predictions.

The first objection, however, can be circumvented in
the case of ``just so" oscillations where the oscillation
length is comparable to one astronomical unit (Glashow and Krauss
1987).
In this case the oscillation probability becomes sharply
energy dependent, and one can choose $\delta m^2$ to
preferentially
suppress one component (e.g., the monochromatic $^7$Be
neutrinos).
This scenario has been explored by several groups and
remains an interesting possibility.  However, the
requirement of large mixing angles remains.

\noindent
\ut{4.2 The Mikheyev-Smirnov-Wolfenstein Mechanism}

     The community's view of neutrino oscillations changed
radically when Mikheyev and Smirnov (1985 and 1986) showed that
the
density dependence of the neutrino effective mass, a phenomenon
first discussed by Wolfenstein (1978), could greatly enhance
oscillation probabilities: a $\nu_e$ is adiabatically transformed
into a $\nu_\mu$ as it traverses a critical density within the
sun.
It became clear that the sun was not only an excellent
neutrino source, but also a natural regenerator for cleverly
enhancing the effects of flavor mixing.

  While the original work of Mikheyev and Smirnov was
numerical, their phenomenon was soon understood analytically
as a level-crossing problem.  If one writes the neutrino
wave function in matter as
\[
|\nu (x)\rangle = a_e (x)|\nu_e\rangle + a_\mu (x)|\nu_\mu\rangle
\]
where $x$ is the coordinate along the neutrino's path, the
evolution of
$a_e(x)$ and $a_\mu(x)$ is governed by
$$i {d \over dx} \left( \matrix { a_{\textstyle e} \cr
a_{\textstyle \mu} \cr} \right) = {1 \over 4E} \left ( \matrix{
2E \sqrt2 G_F \rho(x) - \delta m^2 \cos 2 \theta_{\textstyle v}
{}~~~~~~~~~~~~~~\delta m^2\sin
2\theta_{\textstyle v} \cr
\delta m^2\sin 2 \theta_{\textstyle v} ~~~~~~~~~~~ -2E \sqrt2 G_F
\rho(x) +
\delta m^2
\cos 2\theta_{\textstyle v} \cr} \right) \left( \matrix {
a_{\textstyle e} \cr
a_{\textstyle \mu} \cr} \right) \eqno(20)$$
where G$_F$ is the weak coupling constant and $\rho (x)$ the
solar
electron density.  If $\rho (x)$ = 0, Eq. (20) can be trivially
integrated to give the vacuum oscillation solution (Eq. (17)).
The new contribution to the diagonal elements, $2 E \sqrt2 G_F
\rho(x)$,
represents the effective contribution to $M^2_\nu$  that arises
from neutrino-electron scattering.  The indices of refraction
of electron and muon neutrinos differ because the former
scatter by charged and neutral currents, while the latter
have only neutral current interactions.  The difference in
the forward scattering amplitudes determines the
density-dependent
splitting of the diagonal elements of Eq. (20).

It is helpful to rewrite Eq.(20) in a basis consisting of the
light and heavy
local mass eigenstates (i.e., the states that diagonalize the
right-hand side
of Eq. (20)),
\[
|\nu_L (x)\rangle = \cos \theta (x)|\nu_e\rangle - \sin \theta
(x)|\nu_\mu\rangle
\]
$$
{}~|\nu_H(x)\rangle = \sin \theta (x)|\nu_e\rangle + \cos \theta
(x)|\nu_\mu
\rangle. \eqno(21)
$$

The local mixing angle is defined by
\[
\sin 2 \theta (x)  = {\sin 2 \theta_{\textstyle v} \over
\sqrt{X^2 (x) + \sin^2
2
\theta_{\textstyle v}}}
\]
$$
\cos 2\theta (x)  = {-X (x) \over \sqrt{X^2 (x) + \sin^2
2\theta_{\textstyle
v}}} \eqno(22)
$$
where $X(x) = 2 \sqrt2G_F \rho(x) E/\delta m^2 - \cos
2\theta_{\textstyle v}$.
Thus
$\theta(x)$ ranges from $\theta_{\textstyle v}$ to $\pi/2$ as the
density
$\rho(x)$ goes
from 0 to $\infty$.

If we define
\[
|\nu (x) \rangle = a_H(x)|\nu_H(x)\rangle +
a_L(x)|\nu_L(x)\rangle,
\]
Eq. (20) becomes
$$i {d \over dx} \pmatrix{
a_H \cr
a_L \cr} = \pmatrix {
\lambda(x) & i \alpha (x) \cr
-i \alpha (x) & - \lambda (x) \cr }
\pmatrix
{a_H \cr
a_L }\eqno(23)$$
with the splitting of the local mass eigenstates determined by
$$
2 \lambda (x) = {\delta m^2 \over 2E} \sqrt{X^2 (x) + \sin^2 2
\theta_{\textstyle v}} \eqno(24)
$$
and with mixing of these eigenstates governed by the density
gradient
$$
\alpha (x) = \left({E \over \delta m^2}\right)
 \, {\sqrt2 \, G_F {d \over dx}
\rho(x)
\sin 2 \theta_{\textstyle v} \over X^2 (x) + \sin^2 2
\theta_{\textstyle v}}.
\eqno(25)
$$
Note that the splitting achieves
its minimum value, ${\delta m^2 \over 2E} \sin 2 \theta_v$, at a
critical density $\rho_c =
\rho (x_c)$
$$
2 \sqrt2 E G_F \rho_c = \delta m^2 \cos 2 \theta_v \eqno(26)
$$
that defines the point where the diagonal elements of Eq.~(20)
cross.

Equation (23) can be trivially integrated if the splitting of the
diagonal
elements is
large compared to the off-diagonal elements,
$$
\gamma (x) = \left|{\lambda (x) \over \alpha (x)}\right| =
{\sin^2
2\theta_{\textstyle v} \over \cos
2\theta_{\textstyle v}} \, {\delta m^2 \over 2 E} \, {1 \over |{1
\over \rho_c}
{d \rho (x) \over
dx}|} {[X (x)^2 + \sin^2 2\theta_v]^{3/2} \over \sin^3 2\theta_v}
\gg 1,
\eqno(27)
$$
a condition that becomes particularly stringent near the crossing
point,
$$
\gamma_c = \gamma (x_c) = {\sin^2 2\theta_v \over \cos 2\theta_v}
\, {\delta
m^2 \over 2 E} \, {1 \over \left|{1 \over \rho_c} {d \rho (x)
\over dx}|_{x =
x_c}\right|}
\gg 1. \eqno(28)
$$
The resulting adiabatic electron neutrino survival probability,
valid when
$\gamma_c \gg 1$, is
$$
P^{\rm adiab}_{\nu_e} = {1 \over 2} + {1 \over 2} \cos 2 \theta_v
\cos 2
\theta_i \eqno(29)
$$
where $\theta_i = \theta (x_i)$ is the local mixing angle at the
density where
the neutrino was produced. Eq.~(29) was first discussed by Bethe
(1986)
(also see Messiah 1986).

The physical picture behind this derivation is illustrated
in Figure 12.  One makes the usual assumption that, in vacuum,
the $\nu_e$ is almost identical to the light mass eigenstate,
$\nu_L(0)$, i.e., $m_1 < m_2$ and $\cos \theta_v \sim$ 1.  But as
the density increases,
the matter effects make the $\nu_e$ heavier than the $\nu_\mu$,
with $\nu_e
\to \nu_H (x)$  as $\rho(x)$ becomes large.  The special property
of
the sun is that it produces $\nu_e$s at high density that then
propagate to
the vacuum where they
are measured.  The adiabatic approximation tells us that if
initially $\nu_e \sim \nu_H (x)$, the neutrino will remain on the
heavy
mass trajectory provided the density changes slowly.
That is, if the solar density gradient is sufficiently gentle,
the neutrino will emerge from the sun as the heavy vacuum
eigenstate, $\nu_H (0) \sim \nu_\mu$.  This guarantees nearly
complete conversion
of $\nu_e$s into $\nu_\mu$s, producing a flux that cannot be
detected
by the Homestake or SAGE/GALLEX detectors.

But this does not explain the curious pattern of partial
flux suppressions of Eq. (2).  The key to this is the behavior
when
$\gamma_c \lsim$ 1.  Eqs. (28) and (29) show that the critical
region
for nonadiabatic behavior occurs in a narrow region (for small
$\theta_v$)
surrounding the crossing point, and that this behavior is
controlled by the derivative of the density.  This suggests an
analytic strategy for handling nonadiabatic crossings: one
can replace the true solar density by a simpler (integrable!)
two-parameter
form that is constrained to reproduce the true density and its
derivative at
the crossing point $x_c$. Two convenient choices are the linear
$(\rho (x) = a
+ bx)$ and exponential $(\rho (x) = ae^{-bx})$ profiles.  As the
density
derivative at $x_c$ governs the nonadiabatic behavior, this
procedure should
provide an accurate description of the hopping probability
between the local
mass eigenstates when the neutrino traverses the crossing point.
The initial
and ending points $x_i$ and $x_f$ for the artificial profile are
then chosen
so that $\rho(x_i)$ is the density where the neutrino was
produced in the
solar core and $\rho(x_f) = 0$ (the solar surface), as
illustrated in Fig. 13.
 Since the adiabatic result (Eq. (29)) depends only on the local
mixing angles
at these points, this choice builds in that limit.  Eq. (20) can
then be integrated
exactly for linear and exponential profiles, with the results
given in terms
of parabolic cylinder and Whittaker functions, respectively.
This treatment,
called the finite Landau-Zener approximation (Haxton 1987; Petcov
1988) has
been used extensively in numerical calculations.

We derive a simpler (``infinite") Landau-Zener approximation
(Landau 1932;
Zener 1932) by observing that the nonadiabatic region is
generally confined to
a narrow region around $x_c$, away from the endpoints $x_i$ and
$x_f$.  We
can then extend the artificial profile to $x = \pm \infty$, as
illustrated by
the dashed lines in Fig. 13.  As the neutrino propagates
adiabatically in the
unphysical region $x < x_i$, the exact soluation in the physical
region can be
recovered by choosing the initial boundary conditions
\[
a_L (- \infty) = - a_\mu (- \infty) = \cos \theta_i e^{- i
\int^{x_i}_{-
\infty} \lambda (x) dx} \nonumber\\
\]
$$
a_H (- \infty) = a_e (- \infty) = \sin \theta_i e^{i
\int^{x_i}_{- \infty}
\lambda (x) dx} \eqno(30)
$$
That is $|\nu (-\infty)\rangle$ will then adiabatically evolve to
$|\nu
(x_i)\rangle = |\nu_e\rangle$ as $x$ goes from $- \infty$ to
$x_i$.  The
unphysical region $x > x_f$ can be handled similarly.

With some algebra a simple generalization of the adiabatic
result emerges that is valid for all $\delta m^2/E$ and
$\theta_v$
$$
P_{\nu_e} = {1 \over 2} + {1 \over 2} \cos 2 \theta_v \cos 2
\theta_i ( 1 -
2P_{\rm {hop}}) \eqno(31)
$$
where P$_{\rm {hop}}$ is the probability of hopping from the
heavy mass
trajectory to the light trajectory on traversing the crossing
point.  For the linear approximation to the density,
$$
P^{\rm {lin}}_{\rm {hop}} = e^{- \pi \gamma_c/2}. \eqno(32)
$$
As it must by our construction, $P_{\nu_e}$ reduces to P$^{\rm
{adiab}}_{\nu_e}$ for $\gamma_c \gg$ 1.  The linear Landau-Zener
asymptotic
hopping probability $P^{\rm {lin}}_{\rm {hop}} = e^{- \pi
\gamma_c/2}$ was
derived by Haxton (1986) and independently by Parke (1986), who
married this
approximation to the adiabatic one to get Eq. (31).  The
exponential
probability was first obtained by Petcov (1988),
$$
P^{\rm {exp}}_{\rm {hop}} = {e^{- \pi \delta (1 - \cos 2
\theta_v)} - e^{-2 \pi
\delta} \over 1 - e^{- 2 \pi \delta}}, \eqno(33)
$$
where $\delta = {\gamma_c \cos 2 \theta_v \over \sin^2 2
\theta_v}$.  Note that
 for small $\theta_v, ~~\delta (1 - \cos 2 \theta_v) \to
\gamma_c/2$ and
$\delta \to \gamma_c /(2\theta_v)^2$, so that Eqs. (32) and (33)
then coincide.
When the crossing becomes nonadiabatic (e.g., $\gamma_c \ll 1$ in
Eq. (32)),
the hopping probability goes to 1, allowing the neutrino to
exit the sun on the light mass trajectory as a $\nu_e$, i.e., no
conversion
occurs.

Thus there are two conditions for strong
conversion of solar neutrinos:  there must be a level
crossing (that is, the solar core density must be sufficient
to render $\nu_e \sim \nu_H (x_i)$  when it is first
produced) and the crossing must be adiabatic.  The first
condition requires that $\delta m^2/E$ not be too large, and the
second $\gamma_c \gsim$ 1.  The combination of these two
constraints,
illustrated in Figure 14, defines a triangle of interesting
parameters in the ${\delta m^2 \over E} - \sin^2 2\theta_v$
plane, as Mikheyev and Smirnov
and others (Rosen and Gelb 1986) found by numerically
integrating Eq. (20).  A remarkable feature of this triangle
is that strong $\nu_e \to \nu_\mu$ conversion can occur for very
small
mixing angles $(\sin^2 2 \theta \sim10^{-3}$), unlike the vacuum
case.

One can envision superimposing on Fig. 14 the spectrum of solar
neutrinos, plotted as a
function of ${\delta m^2 \over E}$ for some choice of $\delta
m^2$.
Since Davis sees \ut{some} solar neutrinos, the solutions must
correspond to the boundaries of the triangle in Figure 14.  The
horizontal
boundary indicates the maximum ${\delta m^2 \over E}$ for which
the sun's
central density is sufficient to cause a level crossing.  If a
spectrum
properly straddles this boundary, we obtain a result consistent
with the
Homestake experiment in which low energy neutrinos (large 1/E)
lie above the
level-crossing boundary (and thus remain $\nu_e$'s), but the
high-energy
neutrinos (small 1/E) fall within the unshaded region where
strong conversion
takes place.  Thus such a solution would mimic nonstandard solar
models in
that only the $^8$B neutrino flux would be strongly suppressed.
The diagonal
boundary separates the adiabatic and nonadiabatic regions.  If
the spectrum
straddles this boundary, we obtain a second solution in which low
energy
neutrinos lie within the conversion region, but the high-energy
neutrinos
(small 1/E) lie below the conversion region and are characterized
by $\gamma
\ll 1$ at the crossing density.  (Of course, the boundary is not
a sharp one,
but is characterized by the exponential of Eq. (32)).  Such a
nonadiabatic
solution is quite distinctive since the flux of  pp neutrinos,
which is
strongly constrained in the standard solar model and in any
steady-state
nonstandard model by the solar luminosity, would now be sharply
reduced.
Finally, one can imagine ``hybrid" solutions where the spectrum
straddles both
 the level-crossing (horizontal)
boundary and the adiabaticity (diagonal) boundary for small
$\theta$,
thereby reducing the $^7$Be neutrino flux more than either the
 pp or $^8$B fluxes.

What are the results of a careful search for MSW solutions
satisfying the Homestake, Kamiokande, and SAGE/GALLEX
constraints?
Figure 15 is a calculation by Hata (1995)
(also see Hata and Langacker 1994) for flavor oscillations that
includes the effects of terrestrial regeneration. (MSW effects
can occur
as the neutrinos pass through the earth.)  The preferred
(in the sense of minimizing the $\chi^2$) solution,
corresponding to a region surrounding $\delta m^2 \sim 6 \cdot
10^{-6} eV^2$
and $\sin^2 2\theta_v \sim 6 \cdot 10^{-3}$, is the hybrid case
described above.  It is commonly
called the small-angle solution.  A second, large-angle solution
exists, corresponding to $\delta m^2 \sim 10^{-5} eV^2$ and
$\sin^2 2 \theta_v
\sim$ 0.6, but this region of Figure 15 has shrunk as the
precision
of the gallium experiments improve.

These solutions can be distinguished by their characteristic
distortions of the solar neutrino spectrum.  The survival
probabilities $P_{\nu_e}^{\rm MSW}$(E) for the small- and
large-angle parameters
given above are shown as a function of E in Figure 16.

The calculations of Figure 14 assume flavor oscillations
into a $\nu_\mu$ or $\nu_\tau$.  This influences the
interpretation of
the Kamiokande experiment, as heavy flavor neutrinos contribute
to elastic scattering.  Another possibility is an oscillation
into a sterile neutrino.  The large-angle solution is then
ruled out by the Kamiokande requirement of a large $\nu_e$
survival
probability (see, for example, Hata 1995; Barger, Deshpande, Pal,
Phillips,
and Whisnant, 1991).  It is also ruled by the neutrino
counting limit from big bang nucleosynthesis (Barbieri and
Dolgov 1991; Enqvist el al. 1990; Shi, Schramm, and Fields 1993).

The MSW mechanism provides a natural explanation for the
pattern of observed solar neutrino fluxes.  While it requires
profound new physics, both massive neutrinos and neutrino mixing
are expected in extended models.  The preferred solutions
correspond to $\delta m^2 \sim 10^{-5}$ eV$^2$, and thus are
consistent with
$m_2 \sim$ few $\cdot 10^{-3}$ eV.  This is a typical $\nu_\tau$
mass in models
where $m_R \sim m_{\rm {GUT}}$.  On the other hand, if it is the
$\nu_\mu$
participating in the oscillation, this gives $m_R \sim 10^{12}$
GeV
and predicts a heavy $\nu_\tau \sim$ 10 eV (Bludman, Kennedy, and
Langacker 1992).  Such
a mass is of great interest cosmologically as it would have
consequences for supernova neutrinos (Fuller et al. 1992; Qian et
al. 1993),
the dark matter problem, and the formation of large-scale
structure.

If the MSW mechanism proves not to be the solution of the solar
neutrino problem, it still will have greatly enhanced the
importance of solar neutrino physics: the existing experiments
have ruled out large regions in the $\delta m^2 - \sin^2
2\theta_v$ plane
(corresponding to nearly complete $\nu_e \to \nu_\mu$ conversion)
that
remain hopelessly beyond the reach of accelerator neutrino
oscillation experiments.

\noindent
\ut{4.3 Other Particle Physics Scenarios}

Several other intriguing particle physics phenomena
could affect the solar neutrino puzzle.  The upper bound
established in the earliest $^{37}$Cl runs was consistent with
the complete absence of solar neutrinos, prompting the
suggestion that the $\nu_e$ might decay before reaching earth
(Bahcall, Cabibbo, and Yahil 1972).  This requires a
neutrino mass and a sufficiently fast decay mode.  The
suggestion is less appealing given the present pattern of
fluxes: if the life time is arranged to allow some neutrinos
to survive, one expects the low-energy neutrino flux to be more
severely
suppressed than the high-energy neutrinos.
There is also the constraint from
supernova SN1987A, where $\bar \nu_e$s successfully traveled 50
kpc.
However, one can still wiggle out of both objections if
the neutrino decay is catalyzed by matter effects
\def\mapright#1{\smash{
\mathop{\to}\limits_{#1}}}
$$
\nu_e \mapright{{\rm {MSW}}}\ \nu_2 \to \, decay \,\, products
\eqno(34)
$$
as Raghavan, He, and Pakvasa suggested (1988).  The spectrum
distortions do not
necessarily mimic the MSW mechanism since the decay
probability can depend on the neutrino energy.

Weakly interacting massive particles (WIMPs) were suggested
(Faulkner and Gilliland 1985; Spergel and Press 1985)
as a simultaneous solution to the solar neutrino and dark matter
problems.  If a heavy neutral particle has a mean free path
(at solar densities) comparable to the solar diameter, it can
lose energy in transit and be captured in the sun's gravitational
field.  Once the sun accumulates a sufficiently dense cloud of
such particles, they contribute to energy transport in the sun,
thus violating the SSM assumption that only radiative transport
is important in the core.  The WIMP can pick up energy by
scattering off faster core nucleons, then lose it by rescattering
in the cooler outer layers of the sun.  The additional transport
lowers the core temperature.  While this suggestion is clever,
a simple lowering of $T_c$ is no longer sufficient to reconcile
the
SSM with experiment.

Perhaps the most interesting possibility, apart from the
MSW mechanism, was stimulated by suggestions that the $^{37}$Cl
signal might be varying with a period comparable to the 11-year
solar cycle.  While the evidence for this has weakened,
the original claims generated renewed interest in neutrino
magnetic moment interactions with the solar magnetic field.

The original suggestions by Cisneros (1971) and by
Okun, Voloshyn, and Vysotsky (1986) envisioned the rotation
$$
\nu_{e_L} \to \nu_{e_R} \eqno(35)
$$
producing a right-handed neutrino with sterile interactions in
the standard model.
With the discovery of the MSW mechanism, it was realized
that matter effects would break the vacuum degeneracy of the
$\nu_{e_L}$
and $\nu_{e_R}$, suppressing the spin precession.  Lim and
Marciano
(1988) pointed out that this difficulty was naturally
circumvented
for
$$
\nu_{e_L} \to \nu_{\mu_R} \eqno(36)
$$
as the different matter interactions of the $\nu_e$ and $\nu_\mu$
can
compensate for the vacuum $\nu_e - \nu_\mu$  mass difference,
producing a
crossing similar to the usual MSW mechanism.  Such spin-flavor
precession can then occur at full strength due to an off-diagonal
(in flavor) magnetic moment.

There has been a great deal of clever work on this
problem (Minakata and Nunokawa 1989; Balantekin, Hatchell, and
Loreti 1990).  A very strong limit
on both diagonal and off-diagonal magnetic moments is imposed
by studies of the red giant cooling process of plasmon decay
into neutrinos
$$
\gamma^* \to \nu_i \bar \nu_j. \eqno(37)
$$
The result is $|\mu_{ij}| \lsim 3 \cdot 10^{-12} \mu_B$, where
$\mu_B$
is an
electron Bohr magneton (Raffaelt 1990).  With this bound,
solar magnetic field strengths of $B_\odot \gsim 10^6 G$ are
needed to
produce interesting effects.  Since the location of the
Lim-Marciano level crossing depends on the neutrino energy,
such fields have to be extensive to affect an appreciable
fraction of the neutrino spectrum.  It is unclear whether
these conditions can occur in the sun.

There are interesting, related phenomena involving the
effects of solar density fluctuations or currents on the
MSW mechanism: one can drive $\nu_e \to \nu_\mu$ oscillations in
the
absence of a level crossing by harmonic density perturbations
(not unlike adiabatic fast passage in nuclear magnetic resonance
experiments) (Haxton and Zhang 1991; Sch\"afer and Koonin 1987;
Krastev and Smirnov 1989).  The effects of ``white noise"
density fluctuations on the MSW mechanism have also been
examined recently (Loreti and Balantekin 1994).  Such
fluctuations generate a flavor analog of stochastic spin
depolarization, a phenomenon familiar in atomic physics.

\noindent
5. NEW EXPERIMENTS

The MSW mechanism has had a particularly strong impact
because it was discovered at a time when new data (SAGE/GALLEX,
Kamiokande, helioseismology) were eliminating many competing
solutions to the solar neutrino problem.  The physics of the MSW
mechanism is both simple
and elegant, which accounts for much of its appeal.  But the
most important attribute of this solution is that it can be
definitively tested.  The favored small-angle solution
produces a distinctive distortion in the solar neutrino
spectrum.  Furthermore, if the oscillation is into another flavor
(rather than a sterile state), the missing neutrinos can be
found through their neutral current interactions.
These tests will be made by two high-statistics, direct-counting
directors now under construction.

\noindent
\ut{5.1 The Sudbury Neutrino Observatory}

A water Cerenkov detector of a different type is under
construction deep (5900 m.w.e.) within the Creighton \#9
nickel mine at Sudbury, Ontario, Canada (Ewan et al. 1987;
Aardsma et al. 1987; Chen 1985).  The central portion of the
detector is an acrylic vessel containing 1 kiloton of heavy
water, D$_2$O.  This is surrounded by five meters of light water
to protect the inner detector from neutrons and gammas.  The
detector is viewed by 9500 20-cm PMTs, providing 56\%
photocathode coverage (Figure 17).

The D$_2$O introduces two new channels.  The charged current
breakup reaction
$$
\nu_e + D \to p + p + e^- \eqno(38)
$$
produces a recoil electron which carries off almost all
of the final-state kinetic energy.  As the Gamow-Teller
strength is concentrated very close to the p+p threshold,
1.44 MeV, the electron and neutrino energies are related
by E$_\nu \sim$ E$_e$ + 1.44 MeV.  Thus, neutrino spectrum
distortions should show up clearly in the measured electron
energy distribution.  As the GT strength in the deuteron is
equivalent
to about one-third of a free neutron, the anticipated counting
rates are high.  For an electron detection threshold of 5 MeV
and a $^8$B neutrino flux equal to 51\% of the BP SSM value,
3300 events will be recorded each year.

A second channel is sensitive equally to neutrinos of
any flavor,
$$
\nu_x + D \to \nu'_x + n + p \eqno(39)
$$
and thus will be crucial in testing whether flavor oscillations
have occurred.  The anticipated event rate is approximately
2000/year in the
BP SSM.  The addition of MgCl$_2$ to the D$_2$O at a
concentration of
0.2-0.3\% allows the neutrons to be observed by
$^{35}$Cl(n,$\gamma$).
The Cerenkov light produced by the showering of the 8.6 MeV
capture $\gamma$
ray will add to the signal from the
charged current reaction.  By operating the detector with
and without salt, the experimenters will separate the
charged and neutral current signals.  The SNO collaboration also
plans
to deploy proportional counters filled with $^3$He to exploit
the neutron-specific charge-exchange reaction $^3$He(n,p)$^3$H.
With such
detectors, SNO will be sensitive to neutral current events
at all times.

The detection of $\sim$ 8 neutrons/day in a kiloton detector
places extraordinary constraints on radiopurity.  For example,
a potentially serious background source is the
photodisintegration
of deuterium by energetic photons from U and Th chains.
The experimental goal is concentrations of $\lsim 10^{-14}$ grams
of U and Th
per gram of D$_2$O.

SNO is scheduled to begin operations in mid-1996.

\noindent
\ut{5.2 Superkamiokande}

Superkamiokande will be a greatly enlarged version of Kamiokande
II/III with improved threshold (5 MeV) and energy and position
resolution (Totsuka 1987 and 1990; Takita 1993).
It is currently under construction in the Kamioka mine at a
depth of 2700 m.w.e. (See Figure 18).

The fiducial volume for detecting solar neutrinos will
be 22 kilotons, compared to 0.68 kilotons in the existing
detector.  This plus the improved threshold will increase
the detection rate for neutrino-electron scattering by a
factor of $\sim$ 90, to 8400/year.  Despite the soft kinematics
of the $\nu_e - e$ reaction, the experimenters believe the high
statistics will allow them to distinguish the spectral
distortions
produced by competing MSW solutions.

Because elastic scattering is sensitive to both $\nu_e$ and
heavy-flavor neutrinos (in the ratio of 7:1), an accurate SNO
determination of the $\nu_e$ spectrum will allow Superkamiokande
to extract the spectrum of $\nu_\mu$s or $\nu_\tau$s.

Superkamiokande construction is scheduled to be completed
in March, 1996.

\noindent
\ut{5.3 Other Future Detectors}

The Borexino collaboration (Raghavan 1991; Campanella 1992) has
proposed a 0.3 kiloton liquid scintillator for installation
in the Gran Sasso Laboratory.  The experimenters hope to
detect $^7$Be neutrinos by $\nu - e$ scattering.  The detection
of very low energy recoil electrons places stringent
constraints on U, Th, K, and other activities in the detector,
e.g., $\lsim 10^{-16}$ g U/g, including a requirement for
continuous
purification.  The experimenters will evaluate
background problems in a test facility now under construction,
and scheduled
to be completed by the end of 1995.
The anticipated counting rate for the full-scale detector is
$\sim$
18,000 $^7$Be neutrino events/year for the BP SSM.

A high-counting-rate twin of the $^{37}$Cl detector utilizing the
reaction
$^{127}$I($\nu_e$,e)$^{127}$Xe has been funded recently and is
under
construction in the Homestake mine (Lande 1993; Haxton 1988b).
With a threshold of 664 keV, the detector
is primarily sensitive to $^7$Be and $^8$B neutrinos.  The
initial Homestake detector will contain 100 tons of iodine
as a solution of NaI.  A smaller version of this detector was
recently used at the LAMPF beamstop to
measure the $^{127}$I cross section for stopped muon decay
$\nu_e$s.
Calibration of the $^{7}$Be neutrino cross section by an
$^{37}$Ar
neutrino source is also planned.  The 100-ton detector is
scheduled to
be completed by the end of 1995, with plans for an expansion to 1
kiloton
afterwards.

A 5-kiloton liquid argon time projection chamber,
ICARUS II, has been proposed for Gran Sasso (Rubia 1985).  In
addition to
$\nu - e$ scattering, the charged current channel
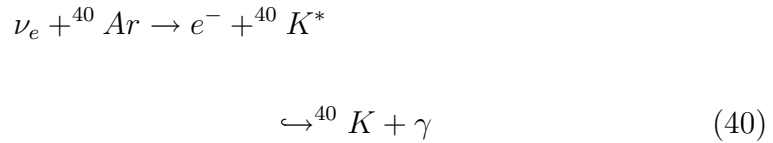
\[
\nu_e +^{40}Ar \to e^-+^{40}K^*
\]
$$
{}~~~~~~~~~~~~~~~~~~~~~~~~~~~~~~~~~~~\hookrightarrow^{40}K+\gamma
\eqno(40)
$$
will allow the experimenters to measure the shape of the
high-energy portion of the $^8$B $\nu_e$ spectrum.

There are a number of important development efforts underway
that focus on new technologies for the next generation
of solar neutrino detectors.
The reader is referred to the recent review by
Lanou (1995).  There are significant challenges
motivating these efforts, e.g., neutrino detection by
coherent scattering off nuclei and real-time detectors
for pp neutrinos, such as the superfluid $^4$He detector HERON
(Bandler et al. 1992) and the high-pressure helium time
projection chamber HELLAZ (Laurenti et al. 1994).

\noindent
6. OUTLOOK

The successes of the Kamiokande and SAGE/GALLEX experiments
have led to a more complicated solar neutrino problem.  The
apparent strong suppression of the $^7$Be flux (negative in
most unconstrained fits!) is not a result expected by those
who favored nonstandard solar model solutions to the $^{37}$Cl
puzzle.  Perhaps this is telling us that solar modelers
have not been sufficiently inventive in modifying the SSM.
But it may also be a push to look elsewhere for the solution.

We do have one candidate solution that works extremely well,
the MSW mechanism.  The required new physics has deep
implications for particle physics and cosmology.  Yet this
physics is not exotic - the requirements of massive neutrinos
and mixing are common assumptions in extensions of the standard
electroweak model.  The elegance of this solution makes it
difficult to maintain one's scientific skepticism: the
notion that the sun was perfectly designed to enhance
the mixing of neutrinos with GUT-scale seesaw masses has
too great an emotional appeal.

Fortunately the solution to the solar neutrino
problem does not have to be decided by community vote; the
issue will be resolved by hard-nosed experimentation.
SNO and Superkamiokande are just a year away, and they
may crack this 30-year-old problem.  Yet these are
difficult experiments, and the physics they address is
fundamental to two of our standard models (particle and solar).
It is prudent
for the community to continue to seek cross checks on these
and other future measurements.  As has proven true in the past,
there is no guarantee that the SNO and Superkamiokande results
will conform to our expectations.

I am indebted to the participants of the Seattle Solar
Modeling Workshop for much of the background material that
was incorporated into this paper.  It is a special pleasure
to thank J. Bahcall, A. B. Balantekin, B. Cleveland, R. Davis
Jr.,
S. Degl'Innocenti, N. Hata, C. Johnson, K. Lande, K. Nakamura, S.
Parke, P. Parker,
Y.-Z. Qian, H. Robertson, Y. Totsuka, and J. Wilkerson for
helpful
discussions and assistance.  This work was supported in
part by the U.S. Department of Energy and by NASA under
grant \#NAGW2523.

\noindent
\ut{Note added in proof (3/24/95)}

Six of the eight planned exposures have been completed in the
SAGE $^{51}$Cr source experiment.  The extractions have
proceeded routinely, and an announcement of the results is
expected later this year.

The BP SSM has been updated by the inclusion of both helium
and heavy element diffusion (Bahcall and Pinsonneault 1995).
The resulting $^8$B neutrino flux is 6.62E6 cm$^{-2}$s$^{-1}$, an
increase
of 16\% (31\%) relative to the BP results with He diffusion
(without diffusion) given in Table 1.  The ``best value"
$^{37}$Cl and $^{71}$Ga capture rates are 9.3 and 137 SNU,
respectively.  The present-day helium surface value is 0.247,
in good agreement with the helioseismology value of 0.242.
The depth of the convective zone is also in excellent agreement
with the value deduced from p-mode oscillation data.
\pagebreak

\noindent
\ut{Literature Cited}

\noindent
Aardsma G, Allan RC, Anglin JD, Bercovitch M, Carter AL, et al.
1987. \ut{Phys.
Lett. B.} \\
\indent
194:321-25

\noindent
Abdurashitov JN, Faizov EL, Gavrin VN, Gusev AO, Kalikov AV, et
al. 1994. \\
\indent
\ut{Phys. Lett. B.} 328:234-48

\noindent
Abraham Z, Iben Jr. I. 1971. \ut{Ap. J.} 170:157-63

\noindent
Adelberger EG, Haxton WC. 1987. \ut{Phys. Rev. C.} 36:879-82

\noindent
Alvarez LW. 1949. \ut{Univ. of California Radiation Lab. Report
UCRL-328},
unpublished

\noindent
Alvarez LW. 1973. \ut{Lawrence Radiation Lab. Physics Notes},
memo \#767

\noindent
Anselmann P, Hampel W, Heusser G, Kiko J, Kirsten T, et al. 1992.
\ut{Phys.
Lett. B.}\\
\indent
285:376-89

\noindent
Anselmann P, Hampel W, Heusser G, Kiko J, Kirsten T, et al. 1994.
\ut{Phys.
Lett. B.}\\
\indent
327:377-85

\noindent
Anselmann P, Frockenbrock R, Hampel W, Heusser G, Kiko J, et al.
1995. \\
\indent
\ut{Phys. Lett. B.} 342:440-50

\noindent
Assenbaum HJ, Langanke K, Rolfs C. 1987. \ut{Z. Physik A.}
327:461-8

\noindent
Austin SM, Anantaraman N, Love WG. 1994. \ut{Phys. Rev. Lett.}
73:30-3

\noindent
Babu KS, Mohapatra RN. 1993. \ut{Phys. Rev. Lett.} 70:2845-48

\noindent
Bahcall JN. 1964a. \ut{Phys. Rev. Lett.} 12:300-2

\noindent
Bahcall JN. 1964b. \ut{Phys. Rev. Lett.} 136:B1164-71

\noindent
Bahcall JN. 1966. \ut{Phys. Rev. Lett.} 17:398-401

\noindent
Bahcall JN. 1989. \ut{Neutrino Astrophysics}. Cambridge:Cambridge
Univ. Press.
567 pp

\noindent
Bahcall JN. 1993. \ut{Phys. Rev. Lett.} 71:2369-71

\noindent
Bahcall JN. 1994. \ut{Phys. Lett. B} 338:276-281

\noindent
Bahcall JN, Bahcall NA, Shaviv G. 1968. \ut{Phys. Rev. Lett.}
20:1209-12

\noindent
Bahcall JN, Bethe H. 1990. \ut{Phys. Rev. Lett.} 65:2233-35

\noindent
Bahcall JN, Cabibbo N, Yahil A. 1972. \ut{Phys. Rev. Lett.}
28:316-8

\noindent
Bahcall JN, Davis Jr. R. 1982. In \ut{Essays in Nuclear
Astrophysics}, ed. CA
Barnes,\\
\indent
DD Clayton, D Schramm, 243.  Cambridge:Cambridge Univ. Press. 562
pp

\noindent
Bahcall JN, Haxton WC. 1989. \ut{Phys. Rev. D.} 40:931-41

\noindent
Bahcall JN, Holstein BR. 1986. \ut{Phys. Rev. C.} 33:2121-7

\noindent
Bahcall JN, Pinsonneault MH. 1992. \ut{Rev. Mod. Phys.}
64:885-926

\noindent
Bahcall JN, Pinsonneault MH. 1995. Submitted to \ut{Rev. Mod.
Phys.}

\noindent
Bahcall JN, Ulrich RK. 1971. \ut{Ap. J.} 170:593-603

\noindent
Bahcall JN, Ulrich RK. 1988. \ut{Rev. Mod. Phys.} 60:297-372

\noindent
Balantekin AB, Bahcall JN, ed. 1995. \ut{Proc. Solar Modeling
Workshop.} Singapore:World\\
\indent
Scientific, in press

\noindent
Balantekin AB, Hatchell PJ, Loreti F. 1990. \ut{Phys. Rev. D.}
41:3588-93

\noindent
Bandler SR, Lanou RE, Maris HJ, More T, Porter FS, Seidel GM,
Torii RH. 1992.\\
\indent
\ut{Phys. Rev. Lett.} 68:2429-32

\noindent
Barbieri R, Dolgov A. 1991. \ut{Nucl. Phys. B.} 349:743-53

\noindent
Barger V, Deshpande N, Pal PB, Phillips RJN, Whisnant K. 1991.
\ut{Phys. Rev.
D.} \\
\indent
43:1759-62

\noindent
Barker FC. 1980. \ut{Aust. J. Phys.} 33:177-90

\noindent
Bartenwerfer D. 1973. \ut{Astron. and Astrop.} 25:455-6

\noindent
Bethe H. 1986. \ut{Phys. Rev. Lett.} 56:1305-8

\noindent
Bludman SA, Kennedy DC, Langacker PG. 1992. \ut{Phys. Rev. D.}
45:1810-13

\noindent
Bowles TJ, Gavrin VN. 1993. \ut{Annu. Rev. Nucl. Part. Sci.}
43:117-64

\noindent
Burbidge EM, Burbidge GR, Fowler WA, Hoyle F. 1957. \ut{Rev. Mod.
Phys.} 29:547

\noindent
Campanella M. 1992. In \ut{Proc. 3rd Int. Workshop on Neutrino
Telescopes},\\
\indent
 ed. M Baldo-Ceolin, p. 73.  Venice: Univ. di Padova

\noindent
Castellani V, Degl'Innocenti S, Fiorentini G, Lissia M, Ricci B.
1994.
\ut{Phys. Rev. D.}\\
\indent
50:4749-61

\noindent
Castellani V, Degl'Innocenti S, Fiorentini G, Ricci B. 1995. See
Balantekin and\\
\indent
Bahcall 1995

\noindent
Charbonnel, C. 1995. See Balantekin and Bahcall 1995

\noindent
Chen HH. 1985. \ut {Phys. Rev. Lett.} 55:534-6

\noindent
Cherry, ML, Fowler WA, Lande K, eds. 1985. \ut{Solar Neutrinos
and Neutrino
Astronomy.}\\
\indent
New York: American Institute of Physics. 322 pp

\noindent
Christensen-Dalsgaard J, Gough DO. 1980. \ut{Nature.} 288:544-7

\noindent
Christy RF, Duck I. 1961. \ut{Nucl. Phys.} 24:89

\noindent
Cisneros A. 1981. \ut{Astrophys. Space Sci.} 10:87

\noindent
Cowan GA, Haxton WC. 1982. \ut{Science.} 216:51-54

\noindent
Davis Jr. R. 1985. See Cherry, Fowler,and Lande 1985, pp 1-21

\noindent
Davis Jr. R. 1993. In \ut{Frontiers of Neutrino Astrophysics},
ed. Y Suzuki
and K Nakamura,\\
\indent
 47. University Academy Press

\noindent
Davis Jr. R, Cleveland BT, Rowley JK. 1983. In \ut{Science
Underground}, ed. MM
Nieto,\\
\indent
 WC Haxton, CM Hoffman, EW Kolb, VD Sandberg, and JW Toevs, 2-15.
\\
\indent
New York, AIP.  446 pp

\noindent
Davis Jr. R, Harmer DS, Hoffman KC. 1968. \ut{Phys. Rev. Lett.}
20:1205-9

\noindent
Demarque P, Mengel J, Sweigart A. 1973. \ut{Ap. J.} 183:997-1004;
\ut{Nature
Phys. Sci.} 246:33

\noindent
Dilke FWW, Gough DO. 1972. \ut{Nature.} 240:262.

\noindent
Dimopoulos S, Hall LJ, and Raby S. 1993. \ut{Phys. Rev. D.}
47:R3697-701

\noindent
Enqvist K, Kainulainen K, Maalampi J. 1990. \ut{Phys. Lett. B.}
249:531-34

\noindent
Ewan GT, Evans HC, Lee HW, Leslie JR, Mak HB, et al. 1987.
Queen's
University\\
\indent
 report SNO-87-12

\noindent
Ezer D and Cameron AGW. 1968. \ut{Astrophys. Lett.} 1:177-9

\noindent
Faulkner J, Gilliland RL. 1985. \ut{Ap. J.} 299:994-1000

\noindent
Filippone BW, Elwyn AJ, Davids CN, Koetke DD. 1983. \ut{Phys.
Rev. Lett.}
50:412-6;\\
\indent
 \ut{Phys. Rev. C.} 28:2222-9

\noindent
Fowler WA. 1972. \ut{Nature.} 238:24

\noindent
Fowler WA. 1984. \ut{Rev. Mod. Phys.} 56:149-79

\noindent
Fuller GM, Mayle RW, Meyer BS, Wilson JR. 1992. \ut{Ap. J.}
389:517-526

\noindent
Gai M, Bertulani CA. 1994. Univ. Conn. preprint 40870-0005

\noindent
Garcia A, Adelberger EG, Mangus PV, Swanson HE, Tengblad HE, et
al. 1991.\\
\indent
\ut{Phys. Rev. Lett.} 67:3654-7

\noindent
Garcia A, Adelberger EG, Mangus PV, Swanson HE, Wells DP, et al.
1995.\\
\indent
\ut{Phys. Rev. C} 51:R439-42

\noindent
Glashow SL, Krauss LM. 1987. \ut{Phys. Lett. B.} 190:199-207

\noindent
Hampel W. 1985. See Cherry, Fowler, and Lande 1985, pp. 162-74

\noindent
Hata N. 1995. See Balantekin and Bahcall 1995

\noindent
Hata N, Haxton WC. 1995. Submitted to \ut{Phys. Lett. B}

\noindent
Hata N, Langacker P. 1994. \ut{Phys. Rev. D.} 48:2937-40

\noindent
Haxton WC. 1986. \ut{Phys. Rev. Lett.} 57:1271-1274

\noindent
Haxton WC. 1987. \ut{Phys. Rev. D.} 35:2352-64

\noindent
Haxton WC. 1988a. \ut{Phys. Rev. C.} 38:2474-77

\noindent
Haxton WC. 1988b. \ut{Phys. Rev. Lett.} 60:768-71

\noindent
Haxton WC, Zhang W-M. 1991. \ut{Phys. Rev. D.} 43:2484-94

\noindent
Hirata KS, Kajita T, Koshiba M, Nakahata M, Oyama Y, et al. 1988.
\ut{Phys.
Rev. D.} \\
\indent
38:448-58

\noindent
Hirata KS, Inoue K, Ishida T, Kajita T, Kihara K, et al. 1991.
\ut{Phys. Rev.
D.} 44:2241-60

\noindent
Iben Jr. I. 1969. \ut{Ann. Phys.} 54:164-203

\noindent
Johnson CW, Kolbe E, Koonin SE, Langanke K. 1992. \ut{Ap. J.}
392:320-7

\noindent
Kavanagh RW, Tombrello TA, Mosher JM, Goosman DR. 1969.
\ut{Bull. Am. Phys. Soc.}\\
\indent
14:1209; Kavanagh RW. 1972. In \ut{Cosmology, Fusion, and Other
Matters},\\
\indent
ed. F Reines. Boulder : Colorado Associated Univ. Press.

\noindent
Krastev PI, Smirnov A Yu. 1989. \ut{Phys. Lett. B.} 226:341-46

\noindent
Krauss A, Becker HW, Trautretter HP, Rolfs C. 1987. \ut{Nucl.
Phys. A.} 467:273-90

\noindent
Krofcheck D, Sugarbaker E, Rapaport J, Wang D, Bahcall JN, et.
al. 1985.\\
\indent
\ut{Phys. Rev. Lett.} 55:1051-4; Krofcheck D. 1987. Ph.D.
thesis,\\
\indent
Ohio State University

\noindent
Kuzmin VA. 1966. \ut{Sov. Phys. JETP}. 22:1051; Kuzmin VA,
Zatsepin GT. 1966.
\\
\indent
\ut{Proc. Int. Conf. on Cosmic Rays} 2:1023

\noindent
Kwong W, Rosen SP. 1994. \ut{Phys. Rev. Lett.} 73:369

\noindent
Landau LD. 1932. \ut{Phys. Z. Sowjetunion}. 2:46

\noindent
Lande K. 1993.  To appear in the \ut{Proc. 22nd Int. Conf. on
Cosmic Ray
Physics}

\noindent
Lande K. 1995. In \ut{Neutrino `94}, ed. A Dar, G Eilam, M
Groneau.\\
\indent
Amsterdam: North Holland

\noindent
Langanke K. 1995. See Balantekin and Bahcall 1995

\noindent
Langanke K, Shoppa TD. 1994. \ut{Phys. Rev. C.} 49:R1771-74

\noindent
Lanou R. 1995. To appear in \ut{Proc. Snowmass Workshop on
Nuclear and
Particle}\\
\indent
\ut{Astrophysics and Cosmology in the Next Millenium}

\noindent
Laurenti G, Tzamarias S, Bonvicini G, Krastev P, Zichichi A et
al. 1994. INFN
preprint

\noindent
Lim CS, Marciano WJ. 1988. \ut{Phys. Rev. D.} 37:1368-73

\noindent
Loreti FN,  Balantekin AB. 1994. \ut{Phys. Rev. D.} 50:4762-70

\noindent
Merryfield WJ. 1995. See Balantekin and Bahcall 1995

\noindent
Merryfield WJ, Toomre J, Gough DO. 1990. \ut{Ap. J.} 353:678-97;
1991. \ut{Ap.
J.} 367:658-65

\noindent
Messiah A. 1986. In \ut{Massive Neutrinos in Astrophysics and in
Particle
Physics},\\
\indent
 ed. O Fackler and J Tran Thanh Van, 373-89.  Gif-sur-Yvette,
Editions
Fronti\`eres. \\
\indent
704 pp

\noindent
Michaud G. 1977. \ut{Nature.} 266:433-4

\noindent
Mikheyev SP, Smirnov A Yu. 1985. \ut{Sov. J. Nucl. Phys.}
42:913-7

\noindent
Mikheyev SP, Smirnov A Yu. 1986. \ut{Nuovo Cimento}. 9C:17-26

\noindent
Mikheyev SP, Smirnov A Yu. 1989. \ut{Prog. in Part. and Nucl.
Phys.} 23:41-136

\noindent
Minakata H, Nunokawa H. 1989. \ut{Phys. Rev. Lett.} 63:121-4

\noindent
Motobayashi T, Iwasa N, Ando Y, Kurokawa M, Murakami H, et al.
1994.\\
\indent
\ut{Phys. Rev. Lett.} 73:2680-3

\noindent
Nakamura K. 1994. To be published in \ut{Proc. Int. Conf. on
Non-Accelerator
Particle}\\
\indent
\ut{Physics} (India)

\noindent
Nico JS. 1995. In \ut{Neutrino `94}, ed. A Dar, G Eilam, M
Groneau.\\
\indent
Amsterdam: North Holland and talk presented at the \ut{Int. Conf.
High}\\
\indent
\ut{Energy Physics}, Glasgow

\noindent
Okun LB, Voloshin MB, Vysotsky MI. 1986. \ut{Sov. Phys. JETP.}
64:446-52 and\\
\indent
\ut{Sov. J. Nucl. Phys.} 44:440-1

\noindent
Opik EJ. 1953. \ut{Contrib. Armagh Obs.} No. 9

\noindent
Parke SJ. 1986. \ut{Phys. Rev. Lett.} 57:1275-8

\noindent
Parke SJ. 1995. \ut{Phys. Rev. Lett.} 74:839-41

\noindent
Parker EN. 1974. \ut{Astrop. Space Sci.} 31:261

\noindent
Parker PD. 1968. \ut{Ap. J.} 153:L85-6

\noindent
Parker PD, Rolfs C. 1991. In \ut{The Solar Interior and
Atmosphere}, ed. A Cox,
\\
\indent
WC Livingston, and MS Mathews, 31.  Tucson:Univ. Arizona Press

\noindent
Petcov ST. 1988. \ut{Phys. Lett. B.} 200:373-9

\noindent
Pontecorvo B. 1946. \ut{Chalk River Report PD-205}, unpublished

\noindent
Pontecorvo B. 1958. \ut{Sov. Phys. JETP.} 7:172

\noindent
Poskanzer AM, McPherson R, Esterlund RA, Reeder PL. 1966.
\ut{Phys. Rev.}
152:995-1001

\noindent
Qian YZ, Fuller FM, Mayle R, Mathews GJ, Wilson JR, Woosley SE.
1993.\\
\indent
\ut{Phys. Rev. Lett.} 71:1965-8

\noindent
Raffaelt G. 1990. \ut{Phys. Rev. Lett.}
 64:2856-8

\noindent
Raghavan RS. 1991. In \ut{Proc. 25th Int. Conf. High Energy
Physics}, ed. KK
Phua and \\
\indent
Y Yamaguchi, 482.  Japan : South Asia Theor. Phys. Assoc. and
Phys. Soc.

\noindent
Raghavan RS, He X-G, Pakvasa S. 1988. \ut{Phys. Rev. D.}
38:1317-20

\noindent
Rhodes Jr. EL, Ulrich RK, Simon GW. 1977. \ut{Ap. J.} 218:901-19

\noindent
Rosen SP and Gelb JM. 1986. \ut{Phys. Rev. D.} 34:969-79

\noindent
Rubia C. 1985. \ut{INFN Publication} INFN/AE-85-7

\noindent
Sch\"afer A, Koonin SE. 1987. \ut{Phys. Lett. B.} 185:417-20

\noindent
Schatzman E, Maeder A. 1981. \ut{Nature.} 290:683-6

\noindent
Sextro RA, Gough RA, Cerny J. 1974. \ut{Nucl. Phys. A.}
234:130-156

\noindent
Shaviv G, Salpeter EE. 1968. \ut{Phys. Rev. Lett.} 21:1602-5.

\noindent
Shi, X, Schramm, D, Fields B. 1993. \ut{Phys. Rev. D.} 48:2563-72

\noindent
Spergel DN, Press WH. 1985. \ut{Ap. J.} 294:663-73

\noindent
Takita M. 1993. In \ut{Frontiers of Neutrino Astrophysics}, ed.
Y. Suzuki and
K. Nakamura,\\
\indent
 135.  Tokyo :  Universal Academic Press.

\noindent
Totsuka Y. 1987.  In \ut{Proc. 7th Workshop on Grand
Unification}, ed. J.
Arafune, 118.\\
\indent
Singapore, World Scientific

\noindent
Totsuka Y. 1990.  In \ut{Proc. Int. Symposium on Underground
Physics
Experiments},\\
\indent
 ed. K. Nakamura, 129.  Univ. of Tokyo

\noindent
Turck-Chi\`eze S, Cahen S, Cass\'e M, Doom C.  1988. \ut{Ap. J.}
335:415-24.

\noindent
Turck-Chi\`eze S, Lopez I. 1993. \ut{Ap. J.} 408:347-67

\noindent
Ulrich RK. 1974. \ut{Ap. J.} 188:369-78

\noindent
Vaughn FJ, Chalmers RA, Kohler D, Chase LF. 1970. \ut{Phys. Rev.
C.} 2:1057-65

\noindent
Walker TP, Steigman G, Schramm DN, Olive KA, Kang H-S. 1991.
\ut{Ap. J.}
376:51-69

\noindent
White M, Krauss L, Gates E. 1993. \ut{Phys. Rev. Lett.} 70:375-8

\noindent
Williams RD, Koonin SE. 1981. \ut{Phys. Rev. C.} 23:2773-4

\noindent
Wolfenstein L. 1978. \ut{Phys. Rev. D.} 17:2369-74; 1979.
\ut{Phys. Rev. D.}
20:2634-35

\noindent
Wolfsberg K, Cowan GA, Bryant EA, Daniels KS, Downey SW, et al.
1995. See
Cherry,\\
\indent
 Fowler, and Lande 1985, pp. 196-202.

\noindent
Xu HM, Gagliardi CA, Tribble RE, Mukhamedzhanov AM, Timofeyuk NK.
1994.\\
\indent
\ut{Phys. Rev. Lett.} 73:2027-30.

\noindent
Zener C. 1932. \ut{Proc. Royal Soc. London.} A137:696

\pagebreak

\noindent
\ut{Figure Captions}

\noindent
Figure 1: The solar  pp chain.

\vspace{.5cm}
\noindent
Figure 2: The flux densities (solid lines) of the principal
$\beta$
decay sources of solar neutrinos of the standard solar model.
The total fluxes
are those of the BP SSM.  The $^7$Be and pep electron capture
neutrino fluxes (dashed lines) are discrete and given in units of
cm$^{-2}$s$^{-1}$.

\vspace{.5cm}
\noindent
Figure 3: The $^7$Be(p, $\gamma) ^8$B S-factors as measured by
Kavanagh
et al. (1969) and by Filippone et al. (1983).  For each data set,
two
theoretical extrapolations, reflecting different choices for the
strong
potentials, to S(0) are shown (Johnson, Kolbe, Koonin, and
Langanke
1992).  The enlargement of the error bars is a correction
by these authors to account for the systematic differences in
the two data sets.

\vspace{.5cm}
\noindent
Figure 4:  The dots represent the $^7$Be and $^8$B fluxes
resulting from the
1000 SSMs of Bahcall and Ulrich (1988), with smaller SSM
uncertainties
added as in Bahcall and Haxton (1989).  The 90 and 99\% c.l.
error
ellipses are shown.

\vspace{.5cm}
\noindent
Figure 5:  The response of the  pp, Be, and $^8$B fluxes to the
indicated
variations in solar model input parameters, displayed as a
function of the
resulting central temperature $T_c$.  From Castellani,
Degl'Innocenti,
Fiorentini, Lissia, and Ricci (1994).

\vspace{.5cm}
\noindent
Figure 6:  The fluxes allowed by the combined results of the
Homestake,
SAGE/GALLEX, and Kamiokande experiments are compared to the
results of SSM
variations and various nonstandard models.  The solid line is the
$T_c$ power
law of Eq. (3).  From Hata (1995).

\vspace{.5cm}
\noindent
Figure 7:  The Homestake $^{37}$Cl solar neutrino experiment.
This schematic
is from the Davis, Harmer, and Hoffman (1968) paper reporting the
first results
from their experiment.

\vspace{.5cm}
\noindent
Figure 8:  Angular distribution of recoil electrons from
Kamiokande II and III showing
the excess at forward angles that is attributed to solar
neutrinos.
Electrons with apparent energies between 7 and 20 MeV are
included.
The upper histogram is the SSM prediction of Bahcall and Ulrich
(1988)
superimposed on an isotropic background, while the lower
histogram is the
best fit.  From K. Nakamura (1994).

\vspace{.5cm}
\noindent
Figure 9:  The Kamiokande II/III recoil electron energy spectrum
compared to the SSM prediction (solid histogram) and to the
SSM prediction rescaled by a factor of 0.5 (dashed).  The last
bin corresponds to electron apparent energies between 14 and 20
MeV.
{}From K. Nakamura (1994).

\vspace{.5cm}
\noindent
Figure 10:  Level scheme for $^{71}$Ge showing the excited states
that
contribute to absorption of  pp, $^7$Be, $^{51}$Cr, and $^8$B
neutrinos.  The
$^{70}$Ge + n break-up threshold is 7.4 MeV.

\vspace{.5cm}
\noindent
Figure 11:  The $^7$Be and $^8$B fluxes determined by the
SAGE/GALLEX and
Kamiokande experiments are compared to the predictions of the BP
and TCL SSMs.
{}From Parke (1995).

\vspace{.5cm}
\noindent
Figure 12:  Schematic illustration of the MSW level crossing.
The dashed lines
correspond to the electron-electron and muon-muon diagonal
elements of the
matrix in Eq. (20).  Their intersection defines the level
crossing density
$\rho_c$.  The solid lines are the trajectories of the light and
heavy local
mass eigenstates.  If the electron is produced deep in the solar
core and propagates adiabatically, it will follow the heavy mass
trajectory,
emerging from the sun as a $\nu_\mu$.

\vspace{.5cm}
\noindent
Figure 13:  The top figure illustrates, for one choice of $\sin^2
\theta_v$ and
$\delta m^2$, that the region of nonadiabatic propagation (solid
line) is
usually confined to a narrow region about the crossing point
$x_c$.  In the
lower figure, the solid lines represent the solar density $\rho
(x)$ and
a linear approximation to $\rho (x)$ that has the correct initial
and final
densities and the correct slope at $x_c$.  (Thus the linear and
exact density
would almost exactly correspond over the nonadiabatic region
illustrated in the
upper figure).  The MSW equation can be solved analytically for
the linear
wedge.  By extending the wedge to $\pm \infty$ (dotted lines) and
assuming
adiabatic propagation in those regions of unphysical density, one
obtains the
simple Landau-Zener result of Eqs. (31) and (32).

\vspace{.5cm}
\noindent
Figure 14:  MSW conversion for a neutrino produced at the sun's
center.  The
upper shaded region indicates those $\delta m^2/E$ where the
vacuum mass
splitting is too great to be overcome by the solar density.  Thus
no level
crossing occurs. The lower shaded region defines the ${\delta m^2
\over E} -
\sin^2 2\theta_v$ region where the level crossing is nonadiabatic
$(\gamma_c <
1)$.  The unshaded region corresponds to adiabatic level
crossings and thus to
strong $\nu_e \to \nu_\mu$ conversion.

\vspace{.5cm}
\noindent
Figure 15:  The MSW solutions allowed at 95\% c.l. by the
combined results of
the Homestake, SAGE/GALLEX, and Kamiokande experiments, including
Kamiokande II
day-night constraints, given the BP flux predictions.  From
Hata (1995).

\vspace{.5cm}
\noindent
Figure 16:  MSW survival probabilities $P^{{\rm MSW}}_{\nu_e}
(E)$ for typical
small-angle $(\delta m^2 \sim 6 \cdot 10^{-6}$eV$^2, \sin^2
2\theta_v \sim 6
\cdot 10^{-3})$ and large-angle $(\delta m^2 \sim 10^{-5}$eV$^2,
\sin^2
2\theta_v \sim 0.6)$ solutions.

\vspace{.5cm}
\noindent
Figure 17:  Schematic of the SNO detector now under construction
in the
Creighton \#9 nickel mine, Sudbury.  Provided by R.G.H. Robertson
and J.F.
Wilkerson (private communication).

\vspace{.5cm}
\noindent
Figure 18:  Photograph of the Kamioka Mine cavity that will house
Superkamiokande.
Photo provided by the Institute for Cosmic Ray Research,
University of Tokyo.

\pagebreak

\setlength{\oddsidemargin}{0in}
\setlength{\evensidemargin}{0in}
\setlength{\textwidth}{7.0in}
\setlength{\topmargin}{-0.25in}
\setlength{\textheight}{8.5in}
\def\ut#1{$\underline{\smash{\vphantom{y}\hbox{#1}}}$}
\renewcommand{\baselinestretch}{2.0}
\def\lsim{\mathrel{\rlap{\lower4pt\hbox{\hskip1pt$\sim$}}
    \raise1pt\hbox{$<$}}}         
\def\gsim{\mathrel{\rlap{\lower4pt\hbox{\hskip1pt$\sim$}}
    \raise1pt\hbox{$>$}}}         
\def\dblint{\mathop{\rlap{\hbox{$\displaystyle\!\int\!\!\!\!\!\in
t$}}
    \hbox{$\bigcirc$}}}
\def\ut#1{$\underline{\smash{\vphantom{y}\hbox{#1}}}$}
\def\overleftrightarrow#1{\vbox{\ialign{##\crcr
    $\leftrightarrow$\crcr
    \noalign{\kern 1pt\nointerlineskip}
    $\hfil\displaystyle{#1}\hfil$\crcr}}}
\long\def\caption#1#2{{\setbox1=\hbox{#1\quad}\hbox{\copy1%
\vtop{\advance\hsize by -\wd1 \noindent #2}}}}
\noindent
Table 1:  Neutrino fluxes predict by the Bahcall/Pinsonneault
(with and without
He diffusion)
and Turck-Chi\`eze/Lopez standard solar models.
$$\vbox {\tabskip 2em plus 3 em minus 1 em
\halign to \hsize {#\hfill && \hfil #\hfil \cr
\noalign {\hrule}
\noalign {\vskip .5pt}
\noalign {\hrule}
\noalign {\vskip 12pt}
Source&
$E_\nu^{\rm max}$ (MeV)&&\ut{flux ($cm^{-2} s^{-1}$)}\cr
&&BP(with diffusion)&BP(without)&TCL&\cr
\noalign {\vskip 12pt}
\noalign {\hrule}
\noalign {\vskip 12pt}
p+p$\to ^2$H + e$^+ + \nu$ &0.42&6.00E10&6.04E10&6.03E10&\cr
$^{13}$N$\to^{13}$C + e$^++\nu$&1.20&4.92E8&4.35E8&3.83E8\cr
$^{15}$O$\to^{15}$N+e$^++\nu$&1.73&4.26E8&3.72E8&3.18E8\cr
$^{17}$F$\to^{17}$O+e$^++\nu$&1.74&5.39E6&4.67E6&\cr
$^8$B$\to^8$Be+$e^++\nu$&$\sim$15&5.69E6&5.06E6&4.43E6\cr
$^3$He+p$\to^4$He+e$^++\nu$&18.77&1.23E3&1.25E3&\cr
$^7$Be+e$^- \to ^7$Li + $\nu$&0.86(90\%)&4.89E9&4.61E9&4.34E9\cr
&0.38(10\%)&&&\cr
p+e$^-$+p$\to^2$H+$\nu$&1.44&1.43E8&1.43E8&1.39E8\cr
\noalign{\vskip 12pt}
\noalign{\hrule}
\noalign{\vskip 12pt}
}}$$

\pagebreak

\setlength{\textwidth}{6.25in}

\noindent
Table 2:  Predicted capture rates in SNU of the BP and TCL SSMs
for the
$^{37}$Cl and SAGE/GALLEX experiments.
$$\vbox {\tabskip 2em plus 3 em minus 1 em
\halign to \hsize {#\hfill && \hfil #\hfil \cr
\noalign {\hrule}
\noalign {\vskip .5pt}
\noalign {\hrule}
\noalign {\vskip 12pt}
&&& capture rates &&\cr
\noalign{\hrule}\cr
neutrino
source&$^{37}$Cl(BP)&$^{37}$Cl(TCL)&$^{71}$Ga(BP)&$^{71}$Ga(TCL)\cr
\noalign {\vskip 12pt}
\noalign {\hrule}
\noalign {\vskip 12pt}
 pp&0.0&0.0&70.8&71.1\cr
pep&0.2&0.22&3.1&2.99\cr
$^7$Be&1.2&1.10&35.8&30.9\cr
$^8$B&6.2&4.63&13.8&10.77\cr
$^{13}$N&0.1&0.063&3.0&2.36\cr
$^{15}$O&0.3&0.21&4.9&3.66\cr
\cr
Total&8.0&6.36&131.5&122.5\cr
\noalign{\vskip 12pt}
\noalign{\hrule}
\noalign{\vskip 12pt}
}}$$

\end{document}